\documentclass[12pt,preprint]{aastex}

\newcommand{\kel}{\rm K}

\newcommand{\rstar}{$R^{\star}$}
\newcommand{\finc}{{\rm F}^{\rm inc}}

\newcommand{\tint}{{\rm T}_{\rm int}}
\newcommand{\teq}{{\rm T}_{\rm eq}}
\newcommand{\teff}{{\rm T}_{\rm eff}}
\newcommand{\phoenix}{{\tt PHOENIX}}
\newcommand{\phx}{{\tt PHOENIX}}

\newcommand{\logg}{\log(g)}
\newcommand{\ang}{\hbox{\AA}}

\newcommand{\tstd}{{\tau}_{std}}

\usepackage{psfig,natbib}

\slugcomment{ApJ, accepted April 2001}

\shortauthors{Barman et al.}
\shorttitle{Irradiated Planets}

\begin{document}
\bibliographystyle{apj}

\title{Irradiated Planets}

       \author{Travis S. Barman, Peter H. Hauschildt}
       \affil{Dept.\ of Physics and Astronomy \& Center for Simulational Physics, 
        University of Georgia, Athens, GA 30602-2451\\
        Email: {\tt travis@hal.physast.uga.edu, yeti@hal.physast.uga.edu}}
       \author{France Allard}
       \affil{C.R.A.L (UML 5574) Ecole Normale Superieure, 69364 Lyon Cedex 7, France\\
        E-Mail: \tt fallard@ens-lyon.fr}

\begin{abstract}
We have modeled irradiated planets located near a dM5 and a G2 primary
star.  The impinging radiation   field was explicitly included in  the
solution of the radiative transfer equation and  in the computation of
the  atmospheric  structure.  We find   that large errors  in both the
thermal  and  reflected flux will   result  from models   which do not
include  the impinging radiation  in a self-consistent manner.  A cool
($\teff   = 500\kel$) and  a hot  ($\teff  =  1000{\kel}$) planet were
modeled at  various orbital separations from both  the dM5  and the G2
primary.  In all scenarios, we compared the effects of the irradiation
in two extreme cases: one where  dust clouds form and remain suspended
in  the atmosphere, and another where  dust clouds form but completely
settle out of the  atmosphere.  The atmospheric structure and emergent
spectrum strongly depend  on the presence or  absence of  dust clouds.
We find that, in the absence of  dust opacity, the impinging radiation
significantly alters the   innermost layers of  an  EGP atmosphere and
that they are  actually brighter  in the  optical than dusty  planets.
Our  models also indicate that  the planet-to-star brightness ratio in
the  optical will be  less that  $1 \times  10^{-5}$  for objects like
$\tau$ b\"ootis which is consistent with recently reported upper  limit
values.
\end{abstract}

\keywords{stars: planetary systems, stars: atmospheres, 
          radiative transfer}

\section{Introduction}

Many indirect  detections of substellar objects in  close orbit around
stars  of   spectral  type   later  than  F   have  been   made  since
1996. However, currently there are few observations that can be useful
for   constraining   the  large   parameter   space  (e.g.    chemical
composition, albedo, age, and inclination)  for any of these so called
extra-solar giant planets (EGPs).  The transits observed for HD209458a
by \cite{Henry00} and \cite{Charbon00} have helped constrain the gross
physical parameters  of HD209458a  (radius, mass, mean  density, etc.)
but not any  of the atmospheric properties.  There  is great hope that
the handful of  ambitious space and ground based  projects planned for
the  next  decade  will   be  capable  of  making  direct  photometric
observations  of  EGPs  and  possibly  measurements  of  EGP  spectral
features.  Until then, we must rely on atmospheric modeling to provide
insight  into the  basic  properties  of these  objects  to guide  the
observers while we wait patiently for their observations.

An early work by \cite{Saumon1996} investigated the properties of EGPs
for various masses and ages near  primaries of different spectral type
but approximated  the reflected and thermal flux   as gray bodies. More
recently there have been several radiative equilibrium models produced
for the  purpose  of predicting certain  observables for  51  Pegasi B
\cite[]{Seager1998} and $\tau$  Boo \cite[]{Gouken2000}. Also, a broad
range (100  --  1700$\kel$) of EGP   models  were studied  and loosely
classified by \cite{Sudarsky2000} using ad hoc temperature -- pressure
profiles.  In this paper,   we present equilibrium models for  several
scenarios  well within the known   parameter space for  EGPs.  We have
investigated the variations of the   thermal structures and   emergent
flux as functions of  both the spectral  type  of the primary  and the
orbital   separation.  We  also  address  the  importance of spherical
versus plane parallel  radiative transfer. Our  emphasis in this paper
is  not   on any  one   particular   EGP,  but instead   on  the basic
understanding   of   these objects  and  the  effects  of  their close
proximity to a stellar companion.

\section{Model Construction}

We have used our multi-purpose atmosphere code \phx\ (version 10.9) to
generate the  models discussed below.   Most details of  the radiative
transfer  method may  be found  in \cite{jcam},  but, for  clarity, we
repeat some of the basic features and discuss a few changes needed for
irradiated models.  \phx\ solves either the full spherically symmetric
or  plane  parallel  radiative  transfer  equation  (PPRTE)  using  an
operator  splitting   (OS)  technique.    For  the  majority   of  the
calculations presented  here, we  have chosen plane  parallel geometry
and  assumed hydrostatic  equilibrium.  However,  we will  discuss the
importance of spherical geometry in section 4.  All models are subject
to  an  energy  conservation  constraint  such  that  the  total  Flux
(convective  and radiative)  is constant  at each  layer.   Each model
atmosphere  spans  a  range  of  optical  depth  ($\tstd$  defined  at
$1.2\micron$) from $0$  at the top of the atmosphere  down to $100$ at
the  deepest layer.  Convection  is treated  according  to the  Mixing
Length  Theory from  the  onset of  the  Schwarzschild criterion  with
mixing length parameter, $\alpha = 1$.

The usual boundary conditions for an isolated star are that the inward
directed    flux   at    the   surface    should   be    zero   (${\rm
I}_{\nu}^{\downarrow}(\tstd=0, \mu)=0$,  where $-1 \le \mu=\cos(\theta)
\le  0$)  and  that the  diffusion
approximation  holds at  the bottom  of the  atmosphere.  For  a close
binary, the  situation is  clearly different.  At  the surface  of the
secondary, the  boundary condition on  ${\rm I}_{\nu}^{\downarrow}$ is
determined by the incident flux (${\rm F}_{\nu}^{\rm inc}$) given by:
\begin{equation}
2\pi\int\limits_{-1}^{0} {\rm I}^{\downarrow}_{\nu}(\mu)\mu d\mu = 
{\rm F}_{\nu}^{\rm inc}(\tstd = 0)
\end{equation}
where
\begin{equation}
{\rm F}_{\nu}^{\rm inc}(\tstd = 0) = 
{\left( \frac{R^\star}{a} \right)}^{2}{\rm F}_{\nu}^{\star} 
\end{equation}
In  the equations above,  ${\rm I}^{\downarrow}_{\nu}(\mu)$  refers to
the inward directed intensities  along direction $\mu$, \rstar\ is the
radius  of   the  primary,   {\it  a}  is   the  surface   to  surface
primary-secondary  separation,  and  ${\rm  F}_{\nu}^{\star}$  is  the
monochromatic flux from the  primary.  For ${\rm F}_{\nu}^{\star}$, we
use  a synthetic  spectrum  taken from  a  previous \phx\  calculation
\cite[]{Nextgen99,Allard2000}.   For the  models  presented below,  we
have  made the  simplifying  assumption that  the impinging  radiation
field is isotropic,  meaning that ${\rm I}^{\downarrow}_{\nu}(\mu)$ at
the   surface    is   the   same   for   all    $\mu$   (i.e.,   ${\rm
I}^{\downarrow}_{\nu}(\mu)={\rm I}^{\downarrow}_{\nu}$).  We have also
assumed that  the flux  is {\it not}  globally redistributed  over the
planet's surface.  A more detailed discussion of these last two issues
may be found in section 4.

As was done in the work  on irradiated M Dwarfs by \cite{Brett93}, all
of the incident  radiation from the primary is  re-radiated outward by
the secondary in the form of reflected flux ($\rm F^{\rm ref}$) and as
a  contribution  to the  thermal  flux  ($\rm  F^{\rm therm}$).   This
constraint  is required by  energy conservation  and implies  that the
integrated flux at the surface  is equal to $\sigma\tint^{4} + \finc$.
Throughout this paper, $\tint$  refers to the effective temperature of
the  planet  in  the   {\it  absence}  of  irradiation  and  $4\pi{\rm
R_{p}}^2\sigma\tint^{4}$  equals  the  planet's  intrinsic  luminosity
where $\rm R_{p}$ is the planet's radius.  The intrinsic luminosity is
an age dependent quantity which  represents the energy released by the
planet  as it cools  and contracts.   $\tint$ also  relates irradiated
planets to isolated planets in  which case $\tint$ is identical to the
more commonly used $\teff$.

The ``effective temperature''  ($\teff$), which is customarily defined
as the temperature a black body would have to have in order to radiate
the total flux, has  important physical and observational significance
for isolated stellar and  substellar objects.  However, for irradiated
planets  (and stars)  $\teff$  loses  some of  its  connection to  the
fundamental  properties  of the  planet  because  it  is difficult  to
separate, by  observation, those photons which  are thermally radiated
by the  planet from  those which originated  from the primary  and are
merely reflected by  the planet.  We shall only  use $\teff$, with its
customary  definition,  to   describe  non-irradiated  objects.   When
describing irradiated  objects, we will define  another quantity which
describes the equilibrium temperature of the planet's day side;
\begin{equation}
\sigma\teq^4 = \sigma\tint^4 + (1-A_{B})\finc,  
\end{equation}
where  $A_B$ is  the Bond  albedo.  Since  the reflected  flux  is not
directly  related   to  the  equilibrium  thermal   structure  of  the
atmosphere,  it  has been  omitted  from  equation  3 and,  therefore,
distinguishes $\teq$  from $\teff$.  It  is also important  to realize
that $\teq$, as  defined above, represents the equilibrium  state at a
given age and allows for the possibility that the intrinsic luminosity
has not  reached zero.  $\tint$ will  only be important  for young (or
more massive) planets when the primary is a solar type star.  However,
for  planets   orbiting  M  dwarfs,  $\tint$  can   be  a  significant
contribution to $\teq$.  While  $\teq$, which likely varies across the
planet's  day  side,  tells  us  little  about  the  interior  or  the
non-irradiated face of the planet  it does provide a useful measure of
the effects on the thermal structure  of the planet's day side.  It is
also  important  to  realize  that  $\tint$  is  not  necessarily  the
equilibrium temperature of the planet's night side since energy may be
carried over from the day side.

To produce an irradiated atmospheric model,  we chose the structure of
a  converged non-irradiated model  taken  from \cite{AMES-2001} as the
initial structure of an irradiated atmosphere located $\sim 5{\rm AU}$
from the primary.  At  this distance, the impinging radiation produces
only small   changes   in  the structure  of   the   upper layers  and
convergence  is achieved after only a  few iterations.  This new model
is then used  as the  initial structure  for an irradiated  atmosphere
located at $\sim   2.5{\rm AU}$ from  the  primary.  This process   of
moving the planet closer to the primary  is repeated until the desired
orbital   separation  is  reached.   Each  intermediate calculation is
iterated until the changes to  the temperature structure are less than
1 $\kel$ at every depth point  and energy conservation is satisfied to
within a  few   percent.   Models obtained  in   this  way  will  have
structures,   at  the    deepest layers,   similar   to  the   initial
non-irradiated model chosen from \cite{AMES-2001},  and thus depend on
$\logg$, and the assumed value of $\tint$.

The  opacity   setups  used  for   the  EGPs  are  identical   to  the
``AMES-Cond''  and  ``AMES-Dusty''  models of  \cite{AMES-2001}  which
refer to two  limiting cases.  AMES-Cond refers to  the situation when
dust forms in  the atmosphere at locations determined  by the chemical
equilibrium  equations,  but  has   been  entirely  removed  from  the
atmosphere  by  efficient  gravitational  settling.   Dust  formation,
therefore,  acts only  to remove  refractory elements  and  reduce the
number  of certain  molecules but  does  {\it not}  contribute to  the
overall  opacity.   The opposite  case,  AMES-Dusty, ignores  settling
altogether.  Dust forms based on the same criteria as in the AMES-Cond
models yet  remains present  to contribute to  the opacity.   The Cond
model may  be thought of  as ``clear skies'',  and the Dusty  model as
``cloudy skies''.

The opacities  include H$_2$O and TiO  lines by \cite{ames-water-new},
CH$_4$  lines from  the HITRAN  and  GEISA databases~\cite[]{hitran92,
geisa},  and H$_2$, N$_2$,  Ar, CH$_4$,  and CO$_2$  collision induced
absorption  (CIA)   opacities  according  to   \cite[][and  references
therein]{h2h2a,  h2hea,  n2n2a,  n2n2b,  h2n2,  n2ch4,  ch4ar,  h2ch4,
ch4ch4,co2co2}.   To   include  dust   grains,  we  have   assumed  an
interstellar size distribution with  diameters ranging from 0.00625 to
0.24$\micron$ and the  chemical equilibrium equations incorporate over
1000   liquids  and  crystals   \cite[]{AMES-2001}.   We   follow  the
prescriptions  of \cite{Grossman72}  and  use Gibbs  free energies  of
formation from the JANAF database \cite[]{janaf} to compute the number
densities  for each  grain species.   The condensation  equilibrium is
directly  incorporated  into  the  chemical equilibrium  equations  to
account, self-consistently,  for the depletion  of refractory elements
as a  function of gas  temperature and pressure.  With  this approach,
any  changes  to the  structure  brought  on  by irradiation  will  be
automatically accounted for in the chemical equilibrium equations.

\section{Results}

We have divided  our study into two  separate cases: one for which the
primary  star    is     a  dM5    ($\teff     = 3000    \kel$,     see
\cite{Leggett2000,Allard2000}) and another  where the primary is  a G2
solar type star ($\teff = 5600 \kel$,  see \cite{Nextgen99}).  For the
former case,  one finds few   observed objects in the  literature  (eg
Gl876B).  However,  surveys     for additional  objects   have   begun
\cite[]{Delfosse1999} .  In the  last decade, most of the observations
have been spent searching for objects that fall into the second group,
and, to date, there are roughly 50 planets known to be in orbit around
solar type stars.    For both of  these cases,  we  consider different
effective  temperatures  and several  orbital  separations.  Below, we
present  the structures  and spectra  for   each case and discuss  the
effects caused by the impinging radiation.

\subsection{EGPs Around M Dwarfs}

Gl876B was  the first planetary  companion found orbiting an  M dwarf.
This was an important discovery  given that M dwarfs constitute nearly
70\% of  all stars in  the galaxy \cite[]{Henry1998}.  Gl876B  is also
one  of the  nearest  EGPs (only  5  pc away)  making  it a  promising
candidate   for  direct   imaging  by   future  adaptive   optics  and
interferometry  missions  \cite[]{Marcy1998}.    Our  first  group  of
irradiated models is intended to represent objects, like Gl876B, which
have cool stellar  primaries.  We begin with a  planet having $\tint =
500\kel$, $\rm R_{\rm  p} = 1\rm R_{\rm jup}$  and $\logg=3.5$ located
very close to a dM5.  Both planet and primary have solar compositions.

In  Figure  \ref{fig1},  we  show  the temperature  structure  of  the
non-irradiated planet compared to irradiated thermal structures at 0.1
and 0.05AU for both the AMES-Cond and AMES-Dusty cases.  The outermost
regions  are  significantly altered  by  the  radiation  from the  dM5
resulting in  a generally  flatter temperature profile.   Without dust
opacity (the  AMES-Cond model), the impinging radiation  is capable of
heating  the   lower  layers  of  the  planet   thereby  reducing  the
temperature gradient.  Above the  photosphere ($\tstd < 10^{-4}$), the
largest  temperature increase  occurs  and a  very slight  temperature
inversion forms  at 0.05AU.  Dust opacity generally  produces a hotter
atmosphere at all depths with a smoother spectral energy distribution.
For more details on the effects of dust in non-irradiated atmospheres,
see \cite{AMES-2001}. It  is important to stress that  the solution of
the chemical  equilibrium equations is based on  the final temperature
structure and thus incorporates the irradiation effects.

In   the   AMES-Dusty  models,   Fe,   Mg$_2$SiO$_4$,  MgSiO$_3$   and
CaMgSi$_2$O$_6$ and MgAl$_2$O$_4$ are  the dominant dust species which
form a cloudy  region extending from roughly $\tstd =  1.0$ to the top
of the atmosphere.  Below this region ($\tstd > 1.0$) is a complicated
mixture  of  various other  condensates.   In  Figure \ref{fig2},  the
concentrations  of  the  dominant  dust  species  are  shown  for  the
irradiated and non-irradiated planets.   There is little change in the
abundance  of the condensates  brought on  by the  impinging radiation
from the dM5.

The spectra for the irradiated and non-irradiated planets are shown in
Figure  \ref{fig3}.  The  isolated AMES-Dusty  planet produces  a very
smooth  spectrum in  the  optical  as is  expected  for an  atmosphere
dominated  by grain opacity.   In the  IR, several  distinct molecular
bands are  present: CH$_4$ at  1.6$\micron$, 2.1 --  2.5$\micron$, and
3.0  -- 4.0$\micron$, and  H$_2$O between  2.5 and  3.0$\micron$.  The
reflective properties of the dust  between the near-UV and the near-IR
is clearly shown in the  spectrum of the irradiated AMES-Dusty planet.
Even at 0.1AU, the planet  reflects a considerable amount of radiation
in the  optical bands compared to the  non-irradiated planet. However,
the ratio ($\epsilon$) of the planet flux to that of the incident flux
is quite small. Averaged over  4500 to 5200$\ang$, $\epsilon$ is about
$10^{-7}$  for the AMES-Dusty  at 0.05AU.   In the  IR, the  flux also
increases and  the molecular  bands become progressively  shallower as
the planet approaches  the primary.  The increase in  flux at infrared
wavelengths is {\it not} due  to reflection, but is entirely a thermal
effect.

The effects of irradiation on  the spectrum of the 500$\kel$ AMES-Cond
planet are also shown in Figure \ref{fig3}.  Distinct molecular bands,
primarily due to water and methane, are clearly visible for $\lambda >
1\micron$.  These  bands are sensitive  to the temperature  at various
depths.   The  methane  absorption   (at  $\sim  1.8\micron$,  2.2  --
3.0$\micron$ and  3.1 -- 4.0  $\micron$) probes the deeper  layers and
indicates  that, up  to 0.1AU,  little change  in the  temperature has
taken  place in  the  photosphere.  Once  0.05AU  separation has  been
reached, the  irradiation has produced a much  larger temperature rise
resulting in  nearly an  order of magnitude  increase in  the emerging
flux  at 3.5$\micron$.   Similarly, the  water bands  (at $0.9\micron,
1.1\micron$ and $1.4\micron$) probe the upper layers and are sensitive
to the irradiation even at 0.1AU.  The AMES-Cond model also produces a
reflected  component  in  the  UV  and optical  regions,  though  less
dramatic than in  the Dusty model when compared  to the non-irradiated
planet's intrinsic flux. In this case, $\epsilon = 3.6 \times 10^{-6}$
at 0.05AU, which is roughly 10 times that seen in the Dusty model.  

It is quite  possible that many of the  young EGPs have $\tint$ closer
to that of an L or T dwarf.   To investigate this possibility, we have
modeled an EGP with  $\tint = 1000\kel$ (all  other parameters are the
same as for  the above  $500\kel$ planet)  irradiated by  the same dM5
used above.   The results  are  qualitatively similar but  weaker, for
both the Cond and Dusty situations, than  those of the previous cooler
planets.  So, in order to explore the regime where irradiation effects
become important, we must study these  planets at even smaller orbital
separations.  Figures  \ref{fig4}    and  \ref{fig5}  demonstrate  the
effects of irradiation on   the structure and  spectrum, respectively,
for  the   1000$\kel$ AMES-Dusty and AMES-Cond    models.  As with the
cooler   dusty planet, the   inner  regions are  not  affected  by the
irradiation even  at  0.005 AU  ($\sim 1\rm R_{\odot}$),  a separation
more common in  cataclysmic variables.  The  AMES-Dusty spectrum (Fig.
\ref{fig5}) is featureless in the IR for separations  $< 0.01 \rm AU$,
but many of the absorption lines present  in the dM5 are now reflected
in  the optical   spectrum  of  the   planet.   The  AMES-Cond  planet
experiences a significant temperature   increase at  all  photospheric
layers (Fig.  \ref{fig4})  and  a suppression  of the  convective zone
also  seen   in  the  cooler AMES-Cond    planet.  The  spectra  (Fig.
\ref{fig5}) show several orders of magnitude  increase in the flux for
the methane and  water  bands  at 0.01AU  with many   reflected atomic
features in the optical when  the Cond planet  is at 0.005AU from  the
primary.

\subsection{EGPs Around Solar Type Stars}

Many EGPs have been discovered orbiting solar type stars (e.g., 51 Peg
B), and have already received much attention in the literature in both
observational and theoretical studies.  Recent works include models of
51  Peg B  \cite[]{Gouken2000}, HD209458a  \cite[]{Seager2000a}  and a
study    of    EGP    photometric    and    polarization    properties
\cite[]{Seager2000b}.  We have taken a slightly more general approach;
we do not  focus on any one particular known  object.  We have chosen,
instead,  a planet  with $\tint  = 500\kel$  and a  young  planet with
$\tint =  1000\kel$ orbiting an object  similar to the  Sun as average
representatives  of most  currently known  EGPs orbiting  F, G,  and K
stars.

Figure \ref{fig6} demonstrates the effects of impinging radiation from
a  G2 primary on  the $500\kel$  AMES-Cond and  AMES-Dusty atmospheric
structure for  several separations.  As  in the dusty  cases presented
above, the  inner part of  the AMES-Dusty atmosphere is  unaffected by
the irradiation.  However, regions  around $\tstd \sim 10^{-4}$ show a
dramatic rise in temperature as the planet is moved from 1.0 to 0.3AU.
At 0.3AU, the Dusty model reaches $\teq = 767\kel$ and the temperature
in the  upper atmosphere  ($\tstd < 10^{-4}$)  has more  than doubled.
Also at 0.3AU,  the AMES-Cond model reaches $\teq  = 735\kel$ and, for
regions near  $\tstd=1$, the temperature has risen  by over 150$\kel$.
Even  the  deepest layers  feel  the  presence  of the  primary.   The
temperature at $\tstd  = 10$ increases by $100\kel$,  and the boundary
between the radiative and convective zone, which was well above $\tstd
=1$  at  1.0AU,   has  retreated  to  nearly  $\tstd   =  50$.   Large
concentrations of  dust species  are still capable  of forming  in the
upper  atmosphere.  In the  AMES-Dusty model,  at 0.15AU,  there still
remains  a thick cloudy  region with  concentrations similar  to those
seen in  the $500\kel$ planet near  a dM5 (see  Fig. \ref{fig2}).  The
only  major difference  being Fe  which extends  to the  upper regions
where $\log({\rm P}_{\rm gas}) \sim 1.0$ dynes cm$^{-2}$.

The optical and IR regions of the spectrum for each orbital separation
are shown  in Figure   \ref{fig7}  where the differences  between  the
AMES-Dusty and AMES-Cond models can clearly be seen.  The dusty models
produce very smooth  spectra except for the  reflected features in the
optical  bands. The dust   is entirely responsible for the  reflection
effects in this  case.  As the dusty  planet is brought closer  to the
G2,     the few absorption  features  (primarily   CH$_4$) seen in the
non-irradiated  planet  completely  disappear.  At 0.5AU,  very little
change has taken  place for $\lambda  >  7000 \ang$ in  the Cond model
spectrum and the water and methane  bands remain strong.  However, for
$\lambda \le 7000 \ang$, a large  amount of reflected light is already
present  even at  0.5AU.  As   the planet is    brought closer to  the
primary, the  reflected  light  around $5000\ang$  steadily increases.
Nearly all ($\sim 95\%$) of  the light reflected by  the Cond model is
due to Rayleigh  scattering by the two  most abundant species, H$_{2}$
and He.   At 0.3AU nearly $10^{7}$ times  more  light emerges from the
planet around $5000\ang$, however   it remains very faint compared  to
the incident radiation with $\epsilon = 2 \times 10^{-7}$ between 4500
and 5200$\ang$.

The younger $\tint  = 1000\kel$ planet behaves in  a similar manner as
in the  cases near a dM5  presented in section 3.1.   The structure of
the  AMES-Cond  planet,  displayed   in  Figure  \ref{fig8},  shows  a
significant suppression of the convective  zone (below $\tstd = 10$ at
0.05AU) and the temperature has increased by a factor of 3 at $\tstd =
10^{-4}$.  The  Dusty model also  shows a large  temperature inversion
above $\log({\rm P}_{\rm gas}) = 3.0$ dynes cm$^{-2}$.  As can be seen
in Figure  \ref{fig9}, the  Dusty model at  0.15AU exhibits  a complex
mixture   of   cloud   species   throughout   the   atmosphere.    Fe,
Mg$_2$SiO$_4$, and MgSiO$_3$ are the most prominent species except for
the  deeper  layers where  CaMgSi$_2$O$_7$  begins  to dominate.   The
spectra, shown  in Figure \ref{fig10},  display similar results  as in
the  previous cases.   However, at  0.05AU, the  molecular  bands have
nearly disappeared  in the Cond  model and the once  broad \ion{Na}{1}
and  \ion{K}{1}  doublets  (5890  and 7680\ang)  are  extremely  weak.
Figure  \ref{fig11} shows,  in more  detail, the  steady  reduction in
equivalent width of both lines for the AMES-Cond models.  The decrease
in equivalent width  is almost entirely due to  the changes in thermal
structure shown above.  At 0.05AU, the Cond planet has reached $\teq =
1752\kel$.

\section{Discussion}

\subsection{Importance of Self-consistent Models}

Many  previous studies have  used the  structures of  a non-irradiated
planets with $\teff = \teq$ and simply computed a spectrum (or albedo)
which  included the  incident flux  and neglected  the effects  on the
structure   and   chemical   composition  \cite[]{Marley1999}.    This
procedure will, however,  result in gross errors in  the emergent flux
from the optical  to infrared.  In Figure \ref{fig12},  we compare the
thermal spectrum of  the self-consistent irradiated (AMES-Cond, $\tint
=1000\kel$)  planet  at   0.065AU  from  a  G2  star   to  that  of  a
non-irradiated AMES-Cond  model with an  equal amount of  thermal flux
(i.e., $\teff = \teq$).  Based  on the irradiated spectrum without the
reflected  component,  we  estimate   that  $\teq  =  1560\kel$.   The
non-irradiated  model  significantly underestimates  the  flux in  the
water  and methane  bands and  overestimates the  flux in  the regions
outside   the  molecular  bands.    Also,  the   non-irradiated  model
overestimates the amount of  reflected light blueward of 6000$\ang$ by
$\sim 35\%$.  Atomic features are also affected as can be seen for the
Na  I  D  line  (Fig.   \ref{fig12}) where  the  equivalent  width  is
overestimated  in the  non-irradiated  model even  when the  reflected
light  is  included  in  the spectrum  calculation.   The  differences
between the  structures (see Fig.  \ref{fig13}) are  just as striking.
As might  be expected,  the non-irradiated model  is still  cooler (by
350$\kel$) than  the irradiated model in the  upper atmosphere ($\tstd
\sim  10^{-4}$).   At  deeper   layers  the  two  structures  actually
intersect and,  at $\tstd \sim  10$, the non-irradiated model  is {\it
hotter} than the irradiated model  by roughly 350$\kel$.  An even more
apparent  difference  is the  location  of  the  boundary between  the
radiative and  convective regions.   In the non-irradiated  model, the
convective  zone reaches  layers  above  $\tstd =  1.0$  while in  the
irradiated  case, the convective  zone has  retreated to  layers below
$\tstd = 10.0$.  These differences, between a non-irradiated structure
with  $\teff  = \teq$  and  a  structure  based on  a  self-consistent
inclusion   of  the   impinging  radiation,   will   have  significant
consequences for any interior and evolution calculations of irradiated
planets.

\subsection{Limiting Effects of Dust}

The question of how much energy is redistributed by "weather patterns"
is extremely  important, and affects the upper  boundary condition for
irradiated models.   To answer this  question, 3D dynamical  models of
grain growth and diffusion would  be needed.  In this work, we explore
the limiting effects of these patterns with Cond models, corresponding
to clear  skies, and Dusty  models corresponding to cloudy  skies.  We
find  that  irradiated  Cond  and Dusty  models  yield  systematically
different  thermal spectra.  The  infrared spectra  of Dusty  and Cond
planets around M  dwarfs are affected by the  impinging radiation only
for extremely small orbits.  However at separations around 0.05AU from
a G2  star, the sudstellar point  (the point on the  planet closest to
the star)  will have  a nearly featureless  spectrum even in  the Cond
limit. The lack of spectral features is a consequence of the structure
becoming  nearly isothermal  and is  not a  result of  any significant
decrease in abundances.

The emergent  flux from both our irradiated  and non-irradiated models
indicates that  Cond atmospheres  are brighter at  optical wavelengths
than  Dusty atmospheres.  In  general, Cond  models are  much brighter
than the  Dusty cases simply because  the dust opacity  blocks most of
the thermal  radiation (see \cite{AMES-2001} for  more details).  When
incident radiation  is present, the dust grains  reflect large amounts
of light while continuing to block most of the {\it intrinsic} optical
flux.  Irradiated  Cond atmospheres remain very  transparent and allow
large amounts  of flux  to emerge from  the deep hotter  layers.  When
this intrinsic thermal radiation  is combined with the reflected flux,
the irradiated Cond model appears brighter than the Dusty model.

The  fraction of the  stellar light  reflected by  our Cond  model (at
0.05AU from a  G2) around 4900$\ang$ is less  than $5 \times 10^{-6}$.
This is well below the results published by \cite{Cameron1999} and the
upper limit  published by \cite{Charbon1999}.  Also,  the strong color
dependence of  Rayleigh scattering causes the  reflected optical light
(scattered  by H$_2$ and  He) in  the Cond  models to  be considerably
different  from the  optical spectrum  of the  primary.   However, the
reflected light in the Dusty atmospheres is due almost entirely to Mie
scattering  which is  a fairly  grey process.   The result  is  a near
reflected copy of the  stellar optical spectrum.  A comparison between
the stellar  light and the reflected  optical spectrum for  a $\tint =
1000$\kel\ planet located  at 0.15AU from a G2 primary  can be seen in
Figure  \ref{fig14}.    For  wavelengths  less   the  $4500\ang$,  the
reflected light in the Dusty  model matches closely the stellar light.
At redder wavelengths, the two spectra differ not only in the slope of
the continuum, but also in  the depth of H$\alpha$ and the \ion{Na}{1}
doublet at  $5890\ang$.  The Cond  model has a very  different optical
spectrum with  the redder wavelengths being dominated  by an extremely
broad \ion{Na}{1}  doublet ($5890\ang$).  In general,  the two spectra
have   little   in   common.    Recent   attempts   have   been   made
\cite[]{Cameron1999,Charbon1999} to observe  a Doppler shifted copy of
the  stellar  light reflected  by  the  planet  orbiting $\tstd$  Boo.
Figure \ref{fig14}  suggests that planets  in the Cond limit  would be
poor  candidates for  such  techniques.  Dusty  planets, however,  are
clearly better choices and  observations at shorter wavelengths may be
more  fruitful.  In  either  case, observations  similar  to those  of
\cite{Cameron1999}  and \cite{Charbon1999}  could  help determine  the
Cond or Dusty nature of EGPs.

\subsection{Energy Redistribution}

Clearly  the fraction  of  the  irradiated face  seen  by an  observer
depends on  orbital phase and inclination.  In  addition, the planet's
atmospheric structure  will likely  vary across the  planet's surface.
\cite{Guillot96}  claimed that  one  could assume  that the  radiation
received by  the planet's day  side is quickly redistributed  over the
entire   planet   surface.    However,   a   recent   calculation   by
\cite{Guillot00}  indicates that  this redistribution  may  take place
over  longer  time-scales  ($\sim  10^{5}$  seconds)  than  previously
thought.  This would  imply that there exists, in  those planets which
are tidally locked, a large temperature difference between the day and
night sides.  Furthermore, if such a difference existed, the structure
of  the irradiated  face  would vary  as  a function  of latitude  and
longitude and should approach the structure of the non-irradiated face
for regions  near the  terminator\footnote{The terminator is  the line
dividing night and day on  a planet.}.  Therefore, it is unlikely that
an  irradiated planet  can  be  characterized by  a  single 1-D  plane
parallel model  atmosphere.  Though it is reassuring  that our results
agree    qualitatively     with    those    of     previous    studies
\cite[]{Seager1998,Seager2000b, Gouken2000}, to accurately predict the
spectrum and  reflected light as  a function of inclination  and phase
will require, at  the very least, many 1-D  models each accounting for
the different amounts of energy  deposited on the planet's day side at
various  longitudes and  latitudes.   We are  currently  working on  a
sequence  of non-isotropic  irradiated models  which include  only the
incident flux  that a  certain latitude\footnote{For this  purpose, we
assume  that the symmetry  axis is  a line  connecting the  planet and
primary.  In this  context, lines of constant latitude  would refer to
concentric rings  about the  substellar point.}  would  receive.  With
such a sequence,  we would be capable of  modeling the reflected light
for any phase and inclination.

Many   of   the   recent  studies   \cite[]{Marley1999,   Seager2000a,
Seager2000b, Gouken2000,  Sudarsky2000} have made  the assumption that
the planet  receives an  amount of energy  equal to $\pi  {\rm R}_{\rm
p}^2 \finc$ which is quickly redistributed \cite[]{Guillot96} over the
entire  planet  surface.   As  a  result, these  previous  works  have
essentially included the minimum amount of external flux ($\frac{1}{4}
\finc$) that could be received by  a planet at a given separation.  If
the more recent calculation  by \cite{Guillot00} is correct and global
redistribution  is  inefficient, then  the  models  in these  previous
studies  would be  valid only  for  separations larger  than what  the
authors had  originally intended and  only for a specific  ring around
the  substellar  point.   In  our  calculations, we  have  assumed  no
redistribution  and therefore have  included a  much larger  amount of
external  flux  ($\finc$).  Despite  the  assumption  of an  isotropic
external radiation field, we feel that our models adequately represent
the substellar point where this assumption is more reasonable.

\subsection{Transmission Spectra}

In the recent study by \cite{Seager2000a} an attempt was made to model
the transmitted  flux of HD209458a  (recently shown to be an eclipsing
system)    through the upper regions of the planet's atmosphere.
\cite{Seager2000a} assumed that the structure at  the poles (and along
the terminator) was the  same as that given  by their fully irradiated
model.  However, the planet  receives only a  small amount of  incident 
flux at the terminator and   only in the   upper atmosphere.  In the
absence  of strong energy redistribution,  the terminator would have a
structure more closely resembling that of the non-irradiated face.

For a transmission study, one  must calculate the stellar flux passing
through a  thin region  at the top  of the planet's  atmosphere ($\sim
0.01{\rm R}_{p}$  thick) which includes  latitudes above $82^{\circ}$.
While,  in general, EGPs  are adequately  described by  plane parallel
geometry, as  one approaches the  limb, the plane  parallel assumption
becomes increasingly less accurate.   Using plane parallel geometry to
calculate  the  transmitted spectrum  assumes  that  the stellar  flux
passes through a region of constant thickness and height (a slab).  In
reality,  the  flux   is  passing  through  a  section   of  a  sphere
encompassing  the limb,  in  which case,  rays  entering at  different
latitudes  will  pass  through   different  amounts  of  the  planet's
atmosphere.   A  simpler and  more  accurate  way  of calculating  the
transmitted  flux  is to  solve  the  spherically symmetric  radiative
transfer  equation (SSRTE)  and  take, as  the  transmitted flux,  the
average  intensities for  direction  cosines which  pass through  this
band.  The main advantage over the plane parallel solution is that the
correct geometry is automatically  built into the SSRTE.  When solving
the  SSRTE,  the  atmosphere  is  modeled  as  a  discrete  number  of
concentric shells  surrounding the  interior (or core).   The transfer
equation is  then solved along characteristics which  are divided into
two  categories: those which  reach the  core (core  intersecting) and
those which  do not (tangential).   Along a given  characteristic, the
direction  cosine is  now a  function of  depth whereas  in  the plane
parallel case it is  constant.  Also, the diffusion approximation must
hold at  the inner boundary for the  core intersecting characteristics
but {\it not} for  the tangential characteristics.  The solution along
the  tangential  characteristics which  pass  through  the outer  most
shells is  all that  is needed to  calculate the transmitted  flux and
will already account for the curvature of the limb.

Consider again  the 1000$\kel$ AMES-Cond planet located  0.05AU from a
G2  primary  and imagine  that  we observe  the  planet  at (or  near)
inferior  conjunction.   Under such  conditions  one  would observe  a
brighter planetary  limb due  to the transmitted  stellar flux.   In a
spherically symmetric geometry, the effect emerges from the model in a
very natural  and physical manner.  In Figure  \ref{fig15}, we compare
the  transmitted   flux  for  two  different   structures:  our  fully
irradiated   AMES-Cond   structure  ($\teq   =   1633\kel$)  and   the
non-irradiated model  ($\teff = 1000\kel$). In both  cases, the planet
leaves its mark on the stellar flux as it passes through the planetary
limb in the form of much stronger absorption features than seen in the
unpolluted  stellar spectrum.  The transmitted  spectrum based  on the
irradiated  structure  peaks  at  $3\micron$, but  the  non-irradiated
planet  (see Fig.  \ref{fig10}) is  still 100  times brighter  in this
region  and over  $10^{5}$ times  fainter than  the primary.   In both
cases, the planet's limb is  $\sim 10^{3}$ times brighter between 4500
and $5000\ang$ than the planet's night side.  However, the disk of the
G2 dominates the optical spectrum  and is $\sim 10^{6}$ times brighter
than the planet's limb.

\section{Conclusions}

In this paper  we have presented atmospheric models  of planets in the
presence  of strong  impinging radiation  from primaries  of different
spectral type and at various orbital separations. We have also studied
the effects  of irradiation in two limiting  cases: efficient settling
(accounting  for  the  depletion  of  elements  by  condensation)  and
complete cloud coverage.

Irradiation has only small effects  on the atmospheres of planets with
dM primaries  except for extremely  close orbits.  For an  object like
Gl876B,  which  orbits  a  dM4  at  0.2AU, it  is  unlikely  that  any
irradiation  effects will be  observed.  However,  for planets  in the
Cond limit orbiting G2 primaries, the effects are non-negligible.  The
upper  layers  of the  atmospheres  are  significantly  heated by  the
impinging radiation even  at 0.5AU in both the  Cond and Dusty limits.
The  inner layers  of  the Cond  models  also experience  considerable
heating and  a suppression of the  radiative-convective boundary.  The
innermost layers of the Dusty models are essentially unaffected by the
irradiation  even for close  orbits.  These  results will  have strong
implications for interior and  evolution calculations.  The heating of
the inner layers  in the Cond limit can bring  the temperatures at the
bottom  of the atmosphere  close to  those in  the Dusty  limit.  This
would suggest  that, in certain situations, the  interior models would
no   longer  depend   on  the   Dust-Cond  uncertainty   published  by
\cite{Chabrier2000}.   However, it  is still  important  that detailed
interior  and  evolution models  be  calculated  which use  irradiated
atmospheres to set upper boundary condition.

The existence of  ``weather patterns'' on  EGPs in the  form of strong
zonal  winds  and clear, cloudy or  partly  cloudy skies, will greatly
influence the effects caused by irradiation.  Our models indicate that
an  EGP with  significant cloud coverage   will reflect a copy of  the
stellar  optical   light while EGPs  with  clear  skies will have very
different reflected spectra.  We have also  found that young EGPs with
clear skies  are  actually brighter  in the optical  than  cloudy EGPs
despite the more efficient reflective  properties of dust. In general,
our models indicate  that observations would need  to be sensitive  to
variations in  the planet+star spectrum on the   order of $10^{-5}$ to
$10^{-6}$ in order to disentangle the reflected or transmitted flux.

It is  apparent from  observations   of Jupiter  and the other  Jovian
planets that clouds are  {\it not} homogeneously distributed  over the
surface and  that complex  weather  patterns persists.   There is also
evidence   (from Galileo)  that, unlike  our  simplified dusty models,
clouds  form in thin  decks  and  any condensation  occurring at   the
uppermost  layers would soon  ``rain'' out  onto the lower  atmosphere
possibly     instigating     a    cascade  of    diffusive    settling
\cite[]{Encrenaz99}.  A self-consistent  treatment of this behavior is
currently being added to \phoenix\ and results for both irradiated and
non-irradiated atmospheres will be presented in  future papers.  Until
these    results  are available,  the     models  presented above  are
representatives  of  clear    (AMES-Cond)   and    completely   cloudy
(AMES-Dusty) skies.  It  is possible  to combine  our  Dusty and  Cond
models to investigate the intermediate ``partly cloudy'' cases.

Though it  is unlikely that  any of the  models produced thus  far are
exact representations of a currently  known EGP, one may hope that our
predictions and those of others  will aid observers in making the much
needed detailed observations of EGPs.

\acknowledgments 

We would like to thank  Tristan Guillot, Adam Showman and  Jean-Pierre
Caillault  for  providing   valuable  comments  and suggestions  which
greatly  improved this paper.  Travis Barman  would especially like to
thank everyone at CRAL,  where a large part  of this work  was carried
out.  Their  hospitality was greatly  appreciated.   This research was
supported by the CNRS and NASA ATP grant  NAG 5-3018, NAG 5-8425, LTSA
grant NAG 5-3619 to the University of Georgia  and LTSA grant NAG5-3435
and NASA EPSCor  grant  to Witchita  State.  Some of  the calculations
presented in this paper were performed on the IBM SP2  at the UGA UCNS
and on the IBM SP2 of the San Diego  Supercomputer Center (SDSC), with
support from the  National Science Foundation, and  on the IBM  SP2 of
the NERSC with support from the DoE.   We thank all these institutions
for a generous allocation of computer time.

\clearpage

\figcaption[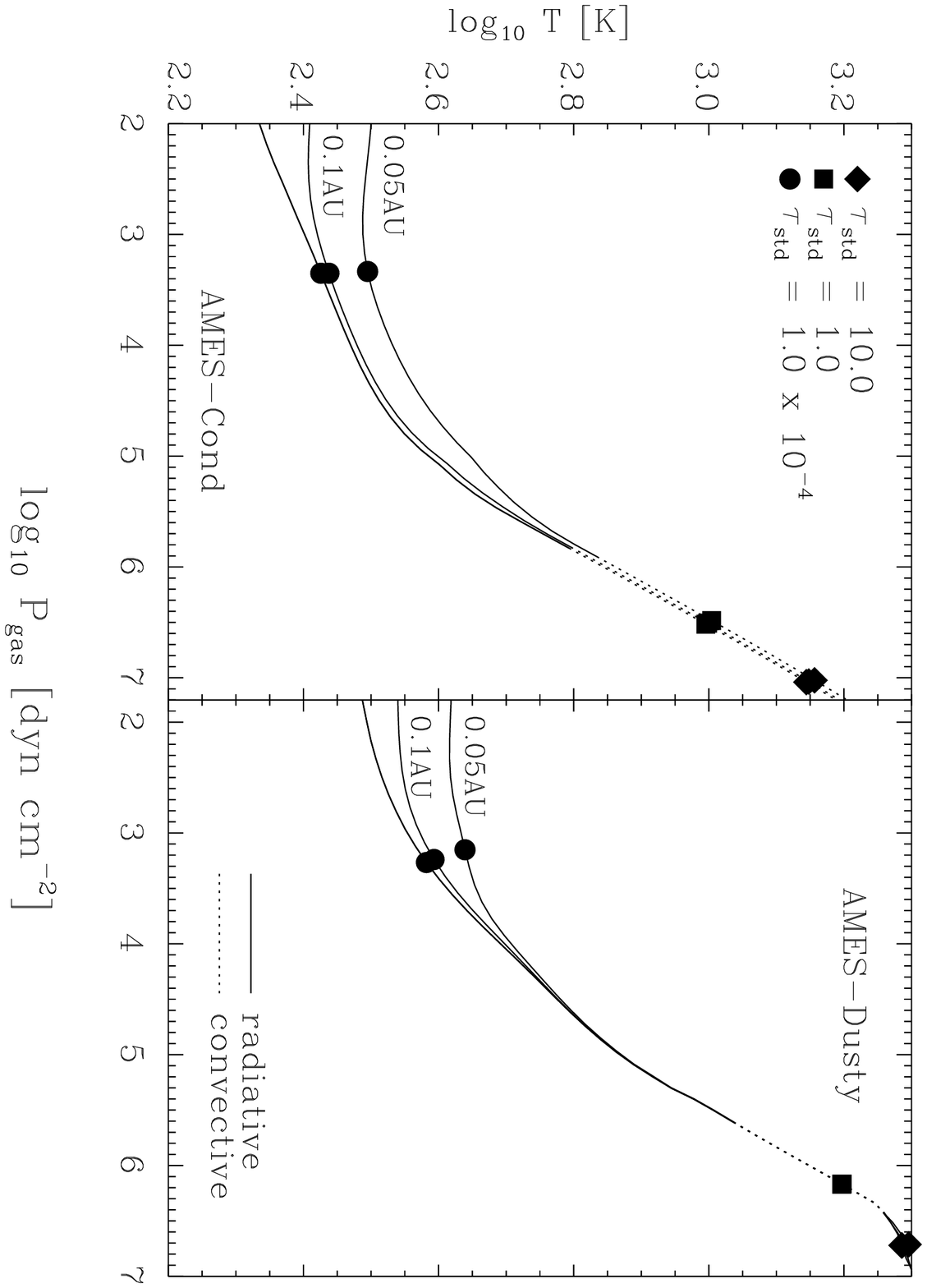]{\label{fig1}  Temperature structures for the non-irradiated
and irradiated planet ($\tint = 500\kel$,  $\logg = 3.5$) when located
0.1 and 0.05AU from a dM5 ($\teff = 3000\kel$).  AMES-Cond is shown on
the left and AMES-Dusty on the right.   The lowest curve in each panel
is the non-irradiated structure. The filled symbols refer to different
optical depths ($\tau$) at $\lambda = 1.2\micron$}

\figcaption[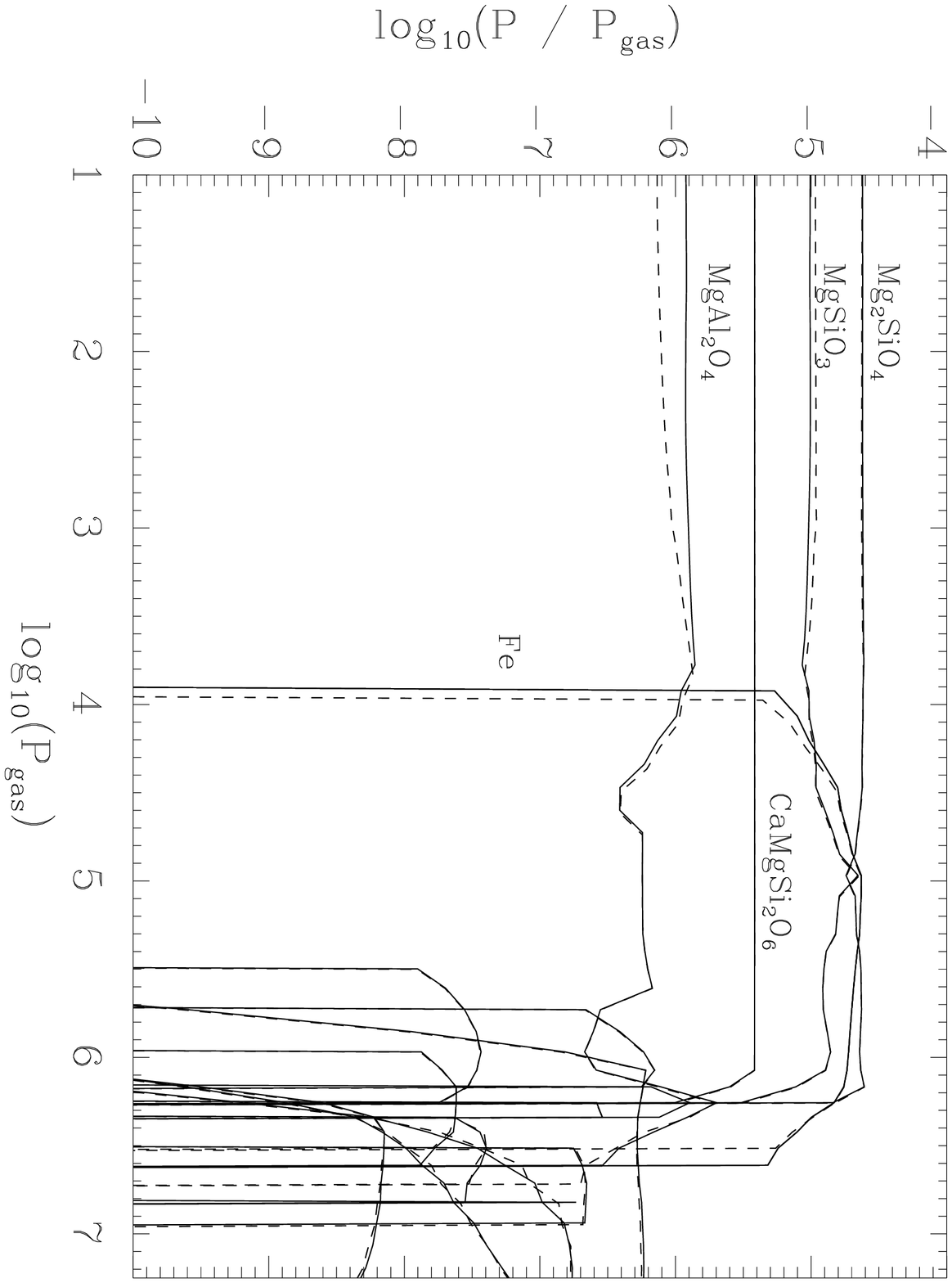]{\label{fig2}  Abundances of  several  important condensates
for  the  irradiated  (solid  lines) AMES-Dusty  ($\tint  =  500\kel$,
$\logg=3.5$) planet  when located 0.05AU from a  dM5.  For comparison,
the non-irradiated abundances are also shown (dashed line).}

\figcaption[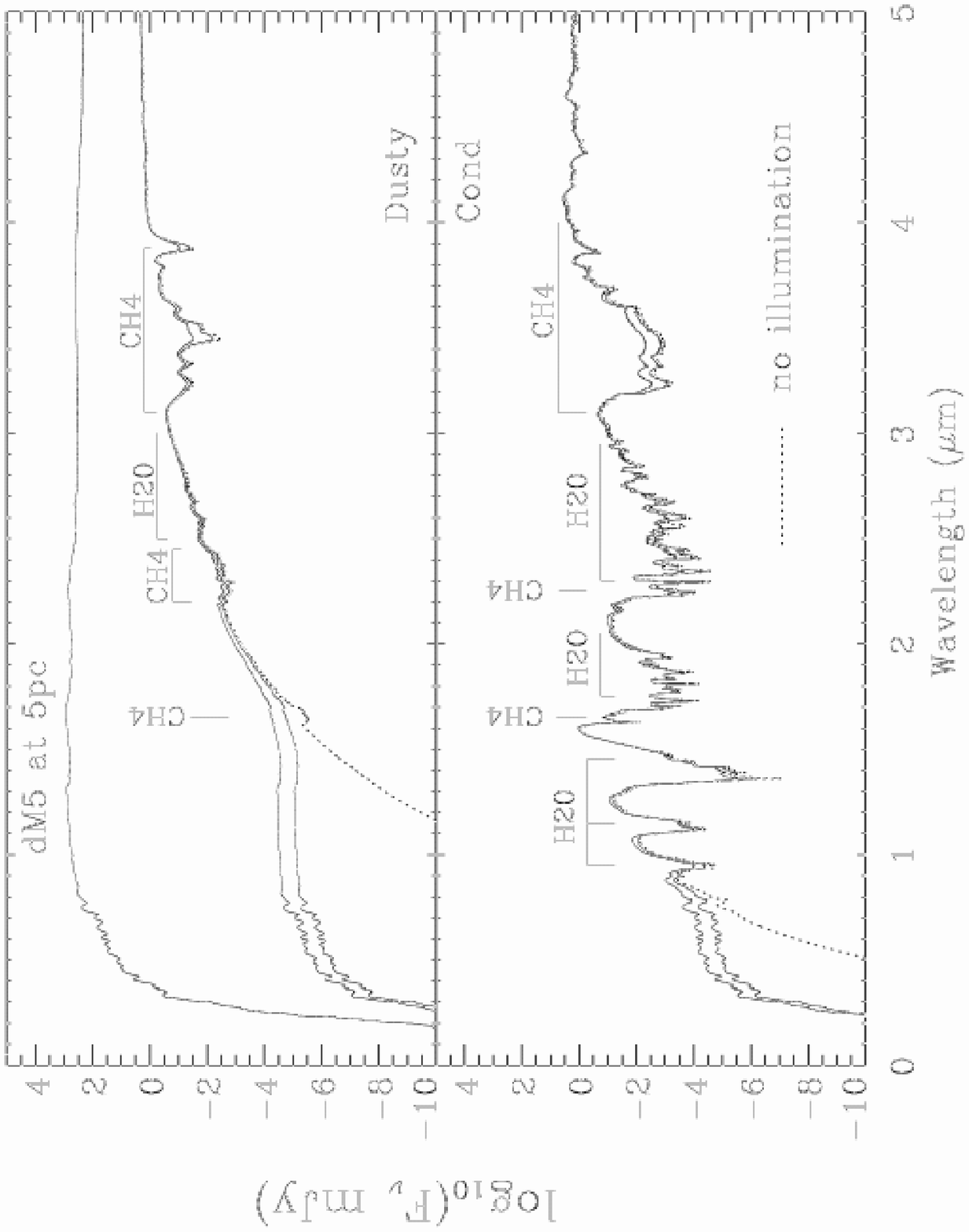]{\label{fig3} Above  are the   spectra corresponding  to the
structures  shown in Fig. \ref{fig1}.  For comparison, the spectrum of
the dM5 used as the  source of irradiation  is also shown.  All fluxes
have been scaled appropriately for the size of the planet (or primary)
and have been  scaled to a distance  of 5 parsecs. AMES-Dusty is shown
on top  and AMES-Cond on the bottom  panel. Note that all spectra have
been heavily smoothed  reducing  the resolution  from $\sim 1\ang$  to
$\sim 50\ang$.}

\figcaption[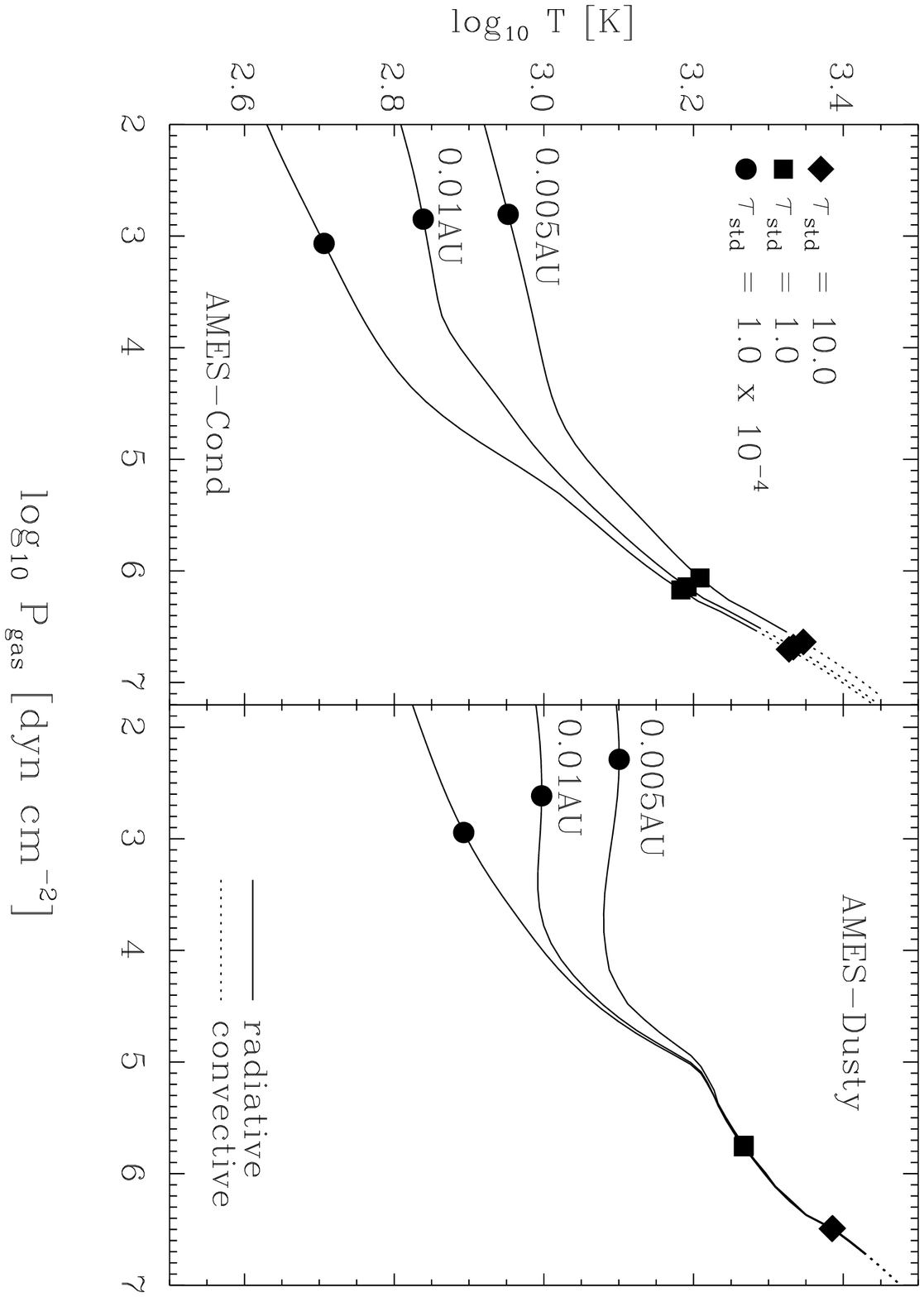]{\label{fig4}  Same as   Fig.  \ref{fig1}, but  now  for the
younger non-irradiated  and   irradiated planet ($\tint   = 1000\kel$,
$\logg=3.5$) when  located 0.01 and 0.005AU  from a dM5.  AMES-Cond is
shown on the left and AMES-Dusty on the right.  As in Fig. \ref{fig1},
the lowest  curve in each panel is  the non-irradiated structure.  The
filled symbols refer to different  optical depths ($\tau$) at $\lambda
= 1.2\micron$. }

\figcaption[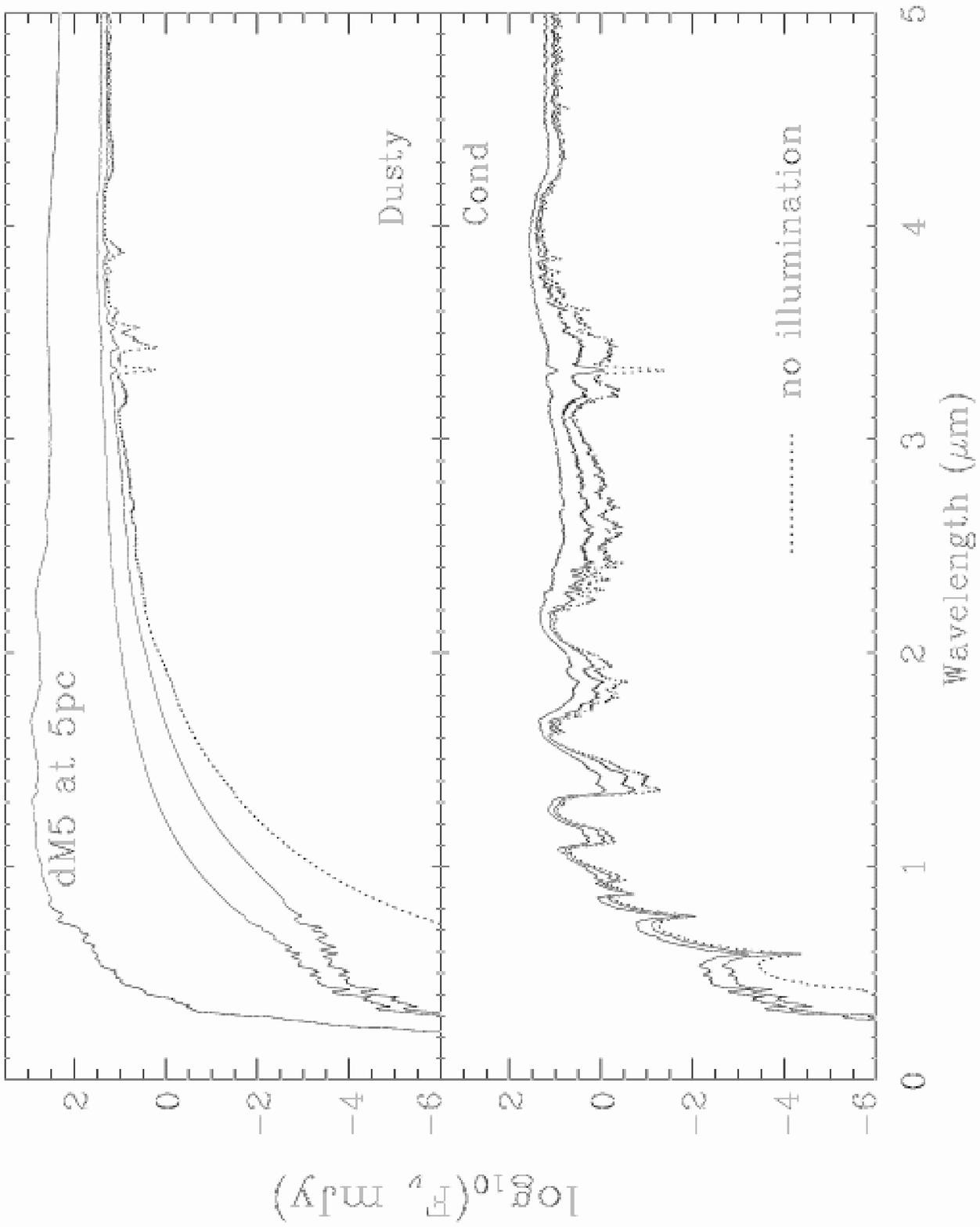]{\label{fig5} Above are the spectra for the structures shown
in Fig.  \ref{fig4}.  For comparison,  the spectrum of the dM5 is also
shown.  All spectra have been scaled as indicated in Fig. \ref{fig3}.}

\figcaption[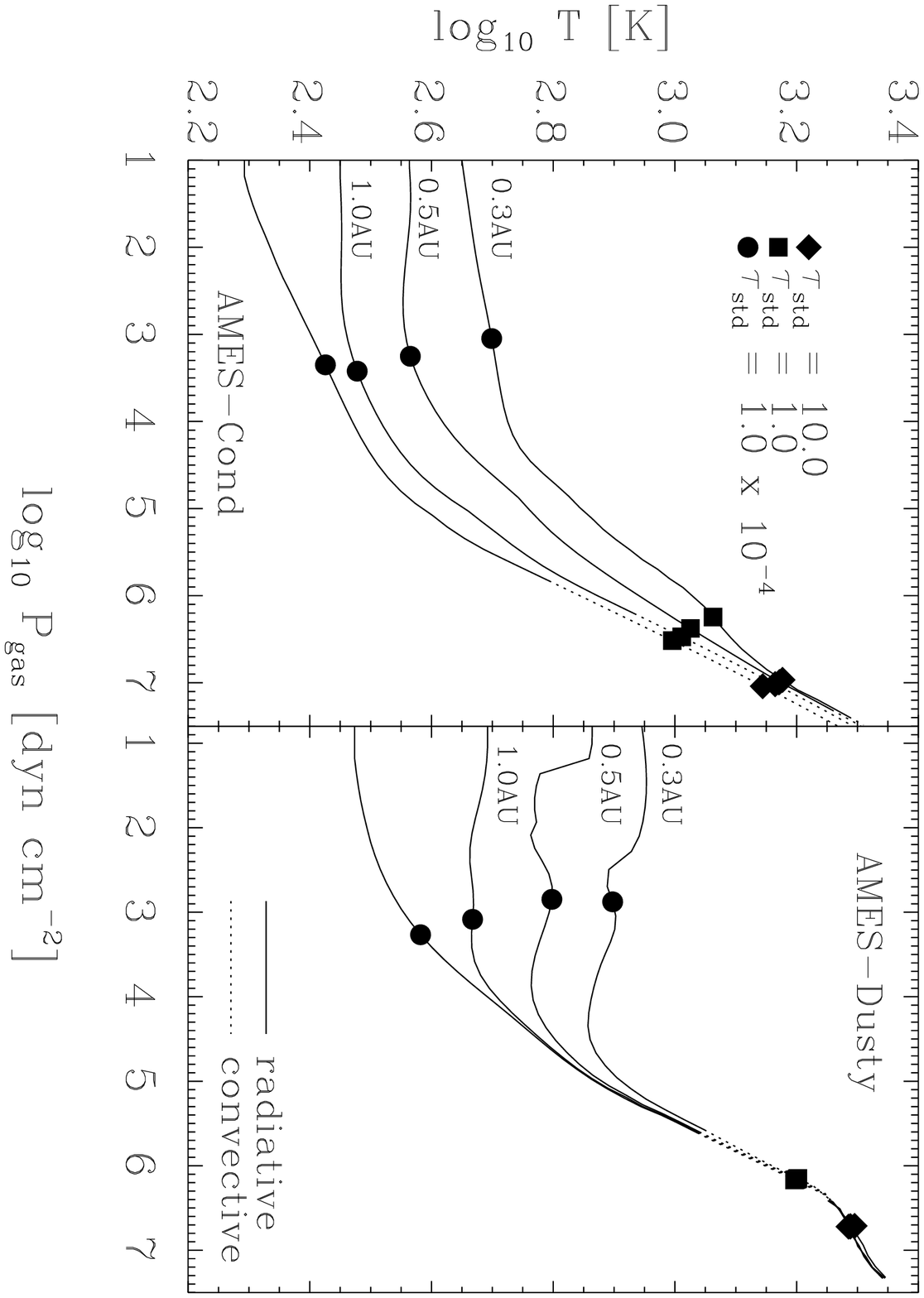]{\label{fig6}  Structures   for  the    non-irradiated   and
irradiated ($\tint =  500\kel$, $\logg=3.5$) planet when  located 1.0,
0.5 and 0.3AU from a G2 primary.The  filled symbols refer to different
optical depths ($\tau$) at $\lambda = 1.2\micron$.}

\figcaption[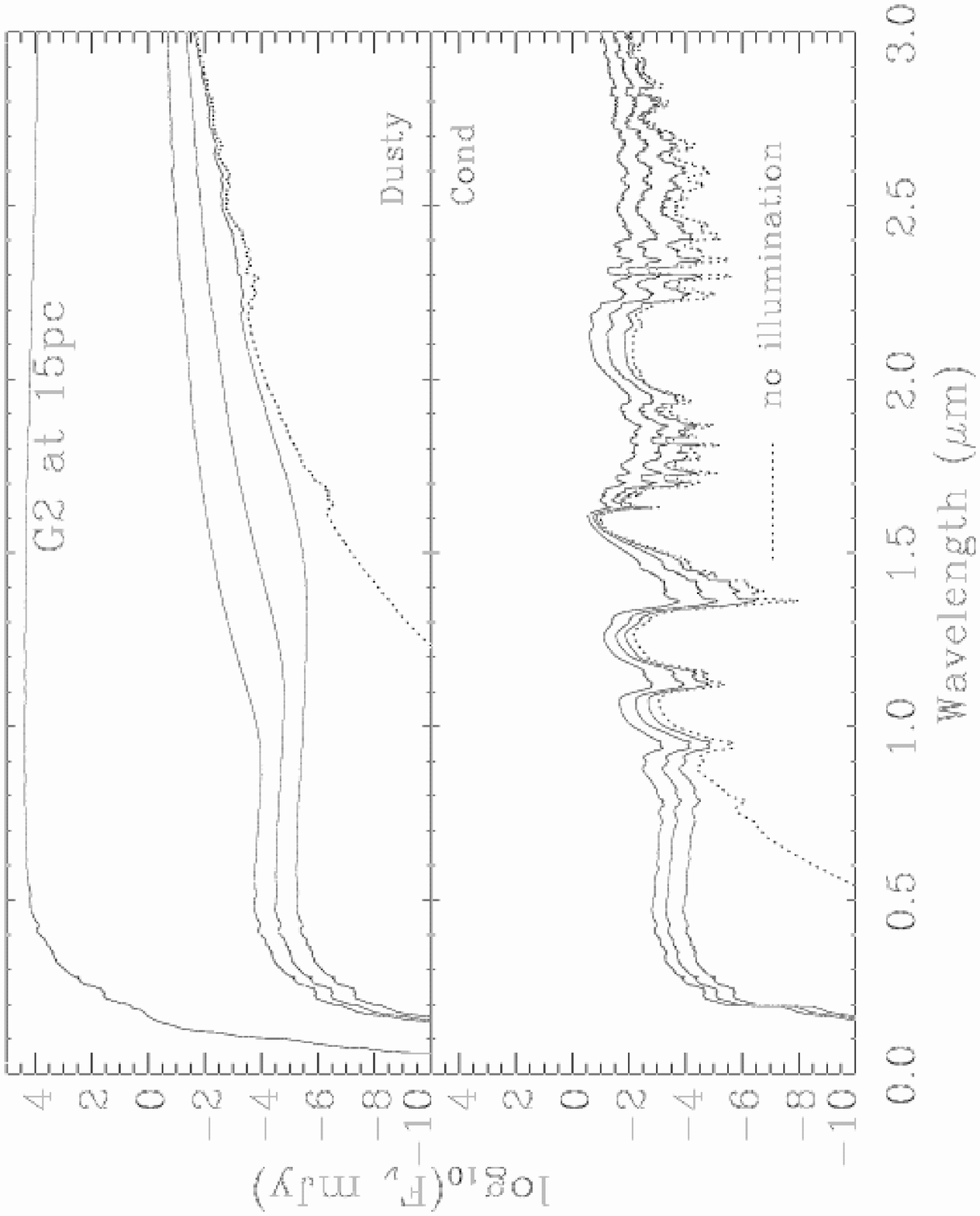]{\label{fig7}  Spectra  for  the  structures shown  in  Fig.
\ref{fig6}.  All spectra have been scaled and smoothed as indicated in
Fig.  \ref{fig3}.  In both panels,  the lowest spectrum  (dotted line)
corresponds to the non-irradiated planet.}

\figcaption[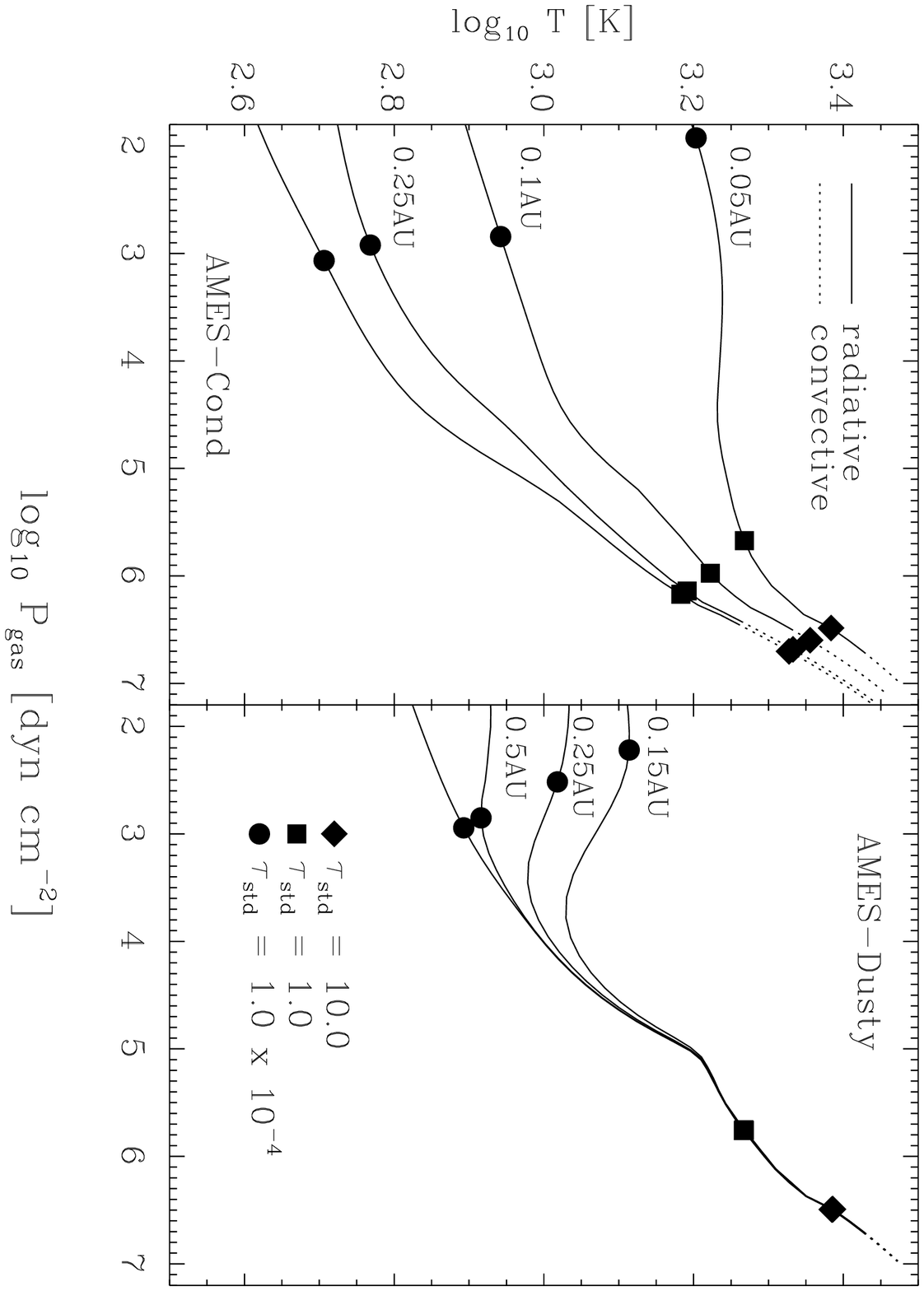]{\label{fig8} Structures    for the     non-irradiated   and
irradiated ($\tint = 1000\kel$, $\logg=3.5$)  planet when located near
a G2 primary. AMES-Cond is shown on the left for 0.25, 0.10 and 0.05AU
separations and  AMES-Dusty is shown on  the right  for 0.5, 0.25, and
0.15AU  separations.   The filled symbols  refer  to different optical
depths ($\tau$) at $\lambda = 1.2\micron$.}

\figcaption[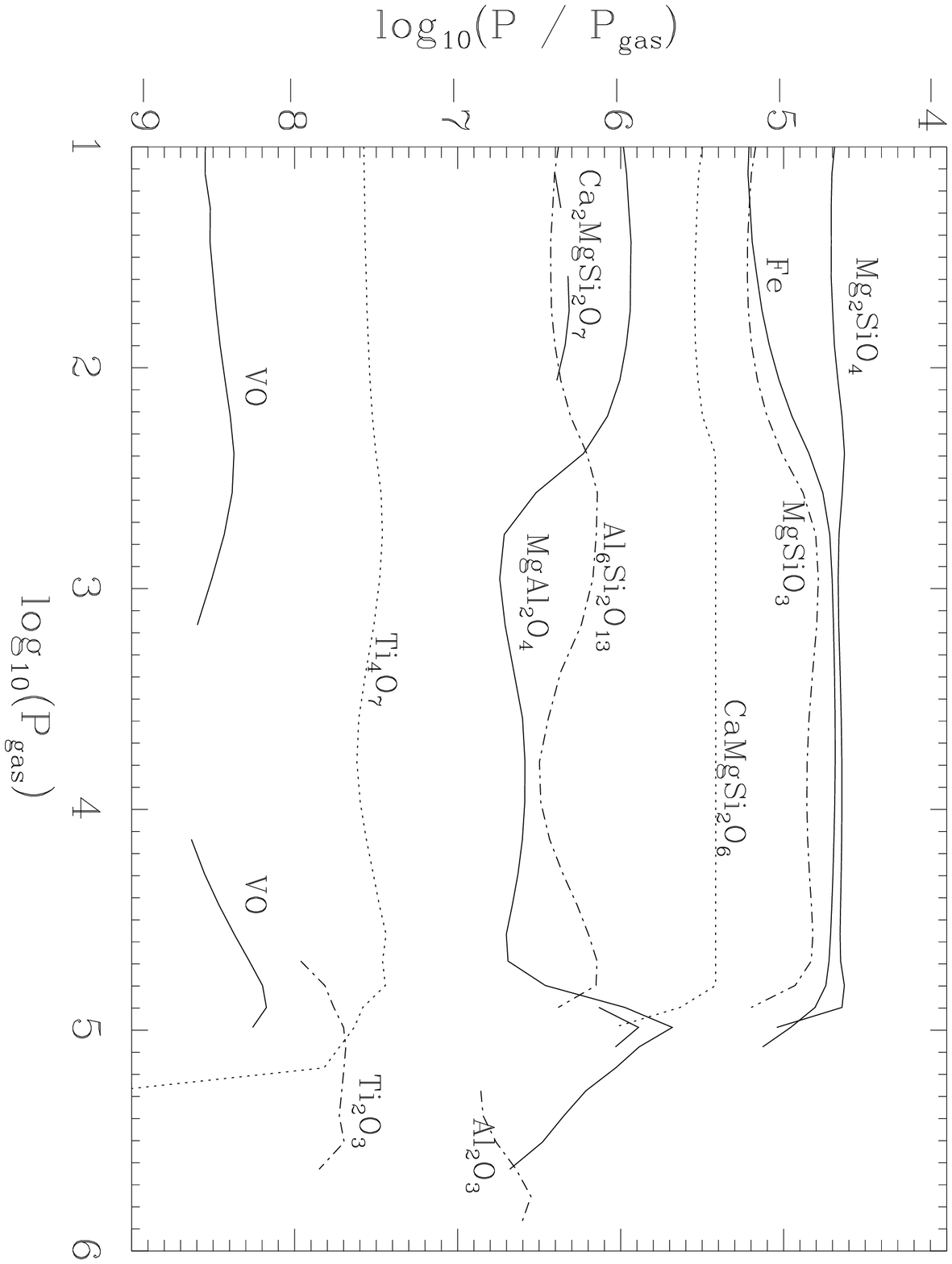]{\label{fig9}Concentrations of several important condensates
for the irradiated AMES-Dusty ($\tint = 1000\kel$, $\logg=3.5$) planet
when located 0.15AU from a G2 primary.}

\figcaption[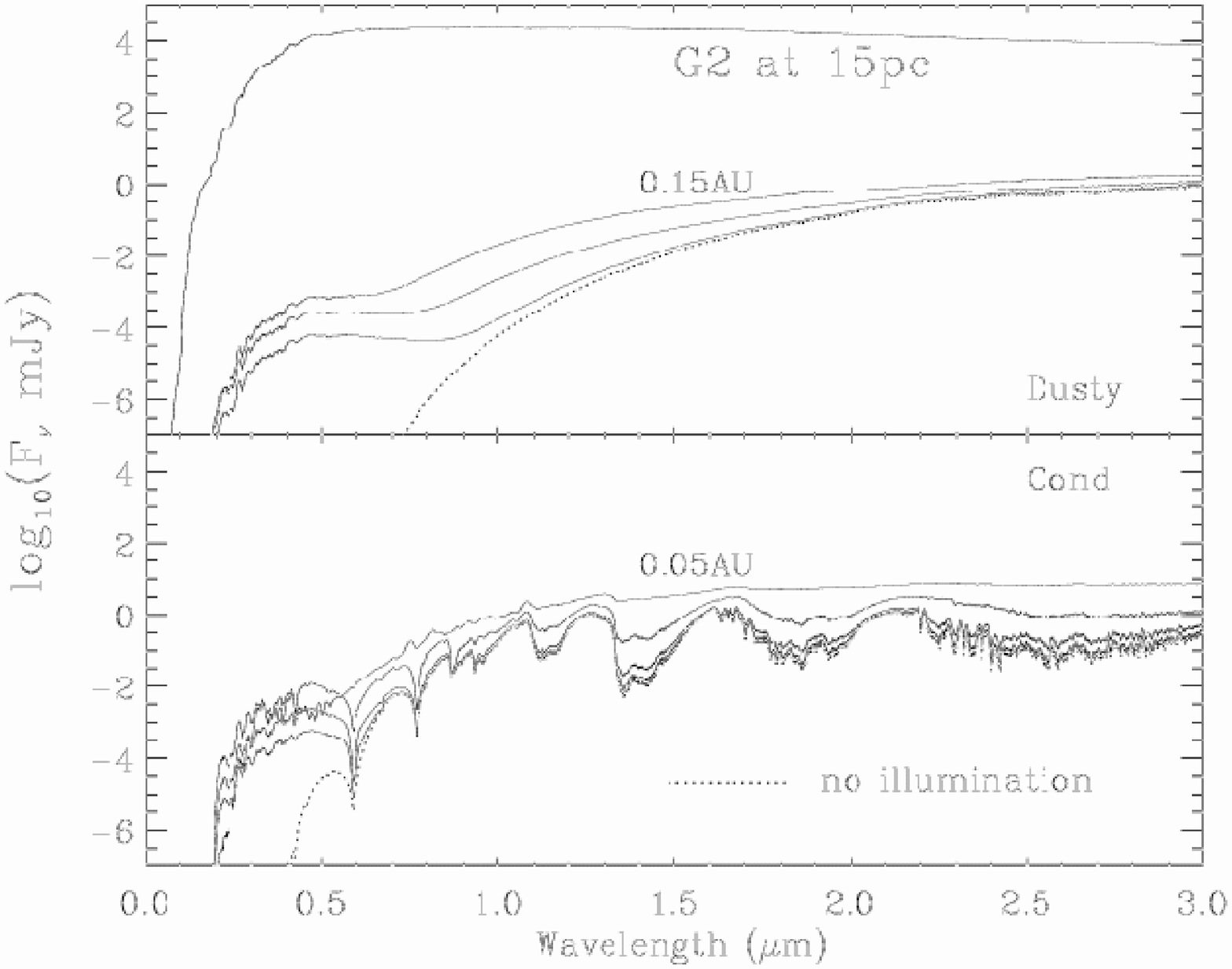]{\label{fig10}  Spectra  for the  structures  shown in  Fig.
\ref{fig9}.  All spectra have been scaled and smoothed as indicated in
Fig.  \ref{fig3}.  In both panels,  the lowest spectrum  (dotted line)
corresponds to the non-irradiated planet.}

\figcaption[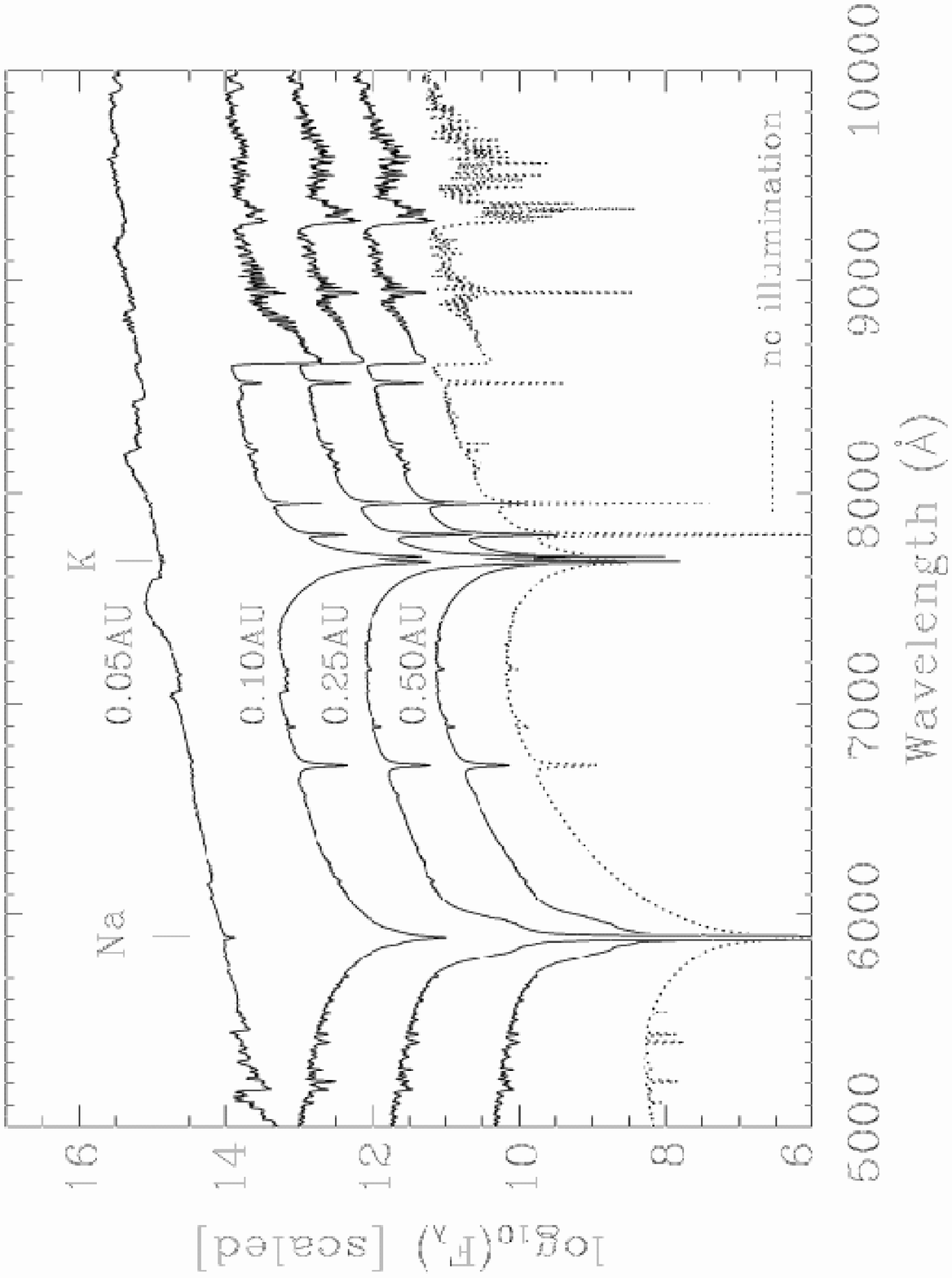]{\label{fig11}Same AMES-Cond  spectra as in
Figure  \ref{fig10} but  focusing  on the  Na  and K  doublet
features. The spectra have been arbitrarily scaled for comparison.}

\figcaption[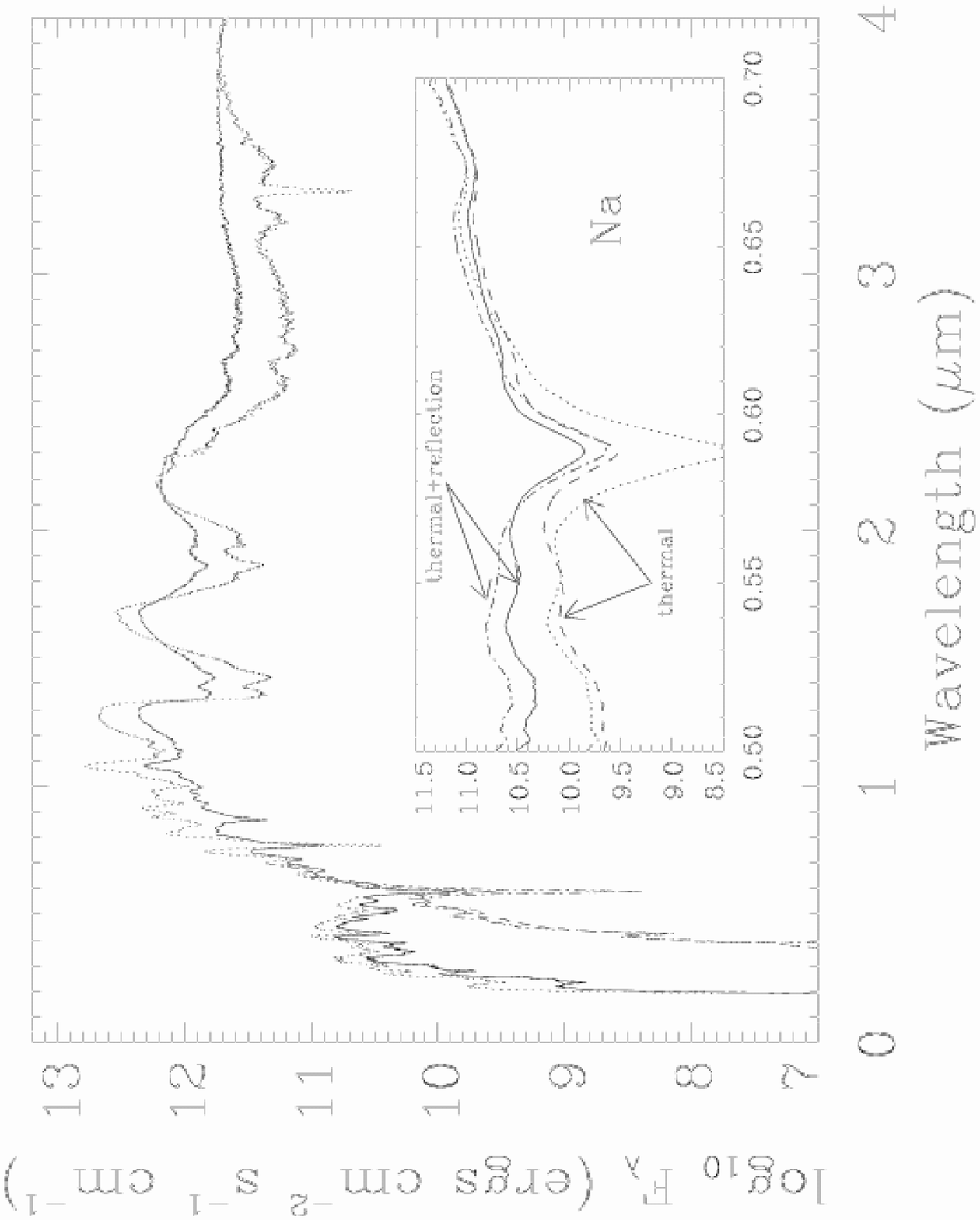]{\label{fig12}  The spectrum  for  our 1000$\kel$  AMES-Cond
planet located at 0.065AU from a solar type primary is shown above with
(solid line) and without (dashed line) the reflected component.  Based
on  the  irradiated  spectrum  without  the  reflected  component,  we
estimate  that $\teq  =  1560\kel$.  We  also  show a  non-irradiated
(AMES-Cond,  $\teff =  \teq$) model  with (dotted  line)  and without
(dashed-dotted  line)  a  reflected  component.   Clearly  the  hotter
non-irradiated  model is  a poor  substitute for  the  true irradiated
case.}

\figcaption[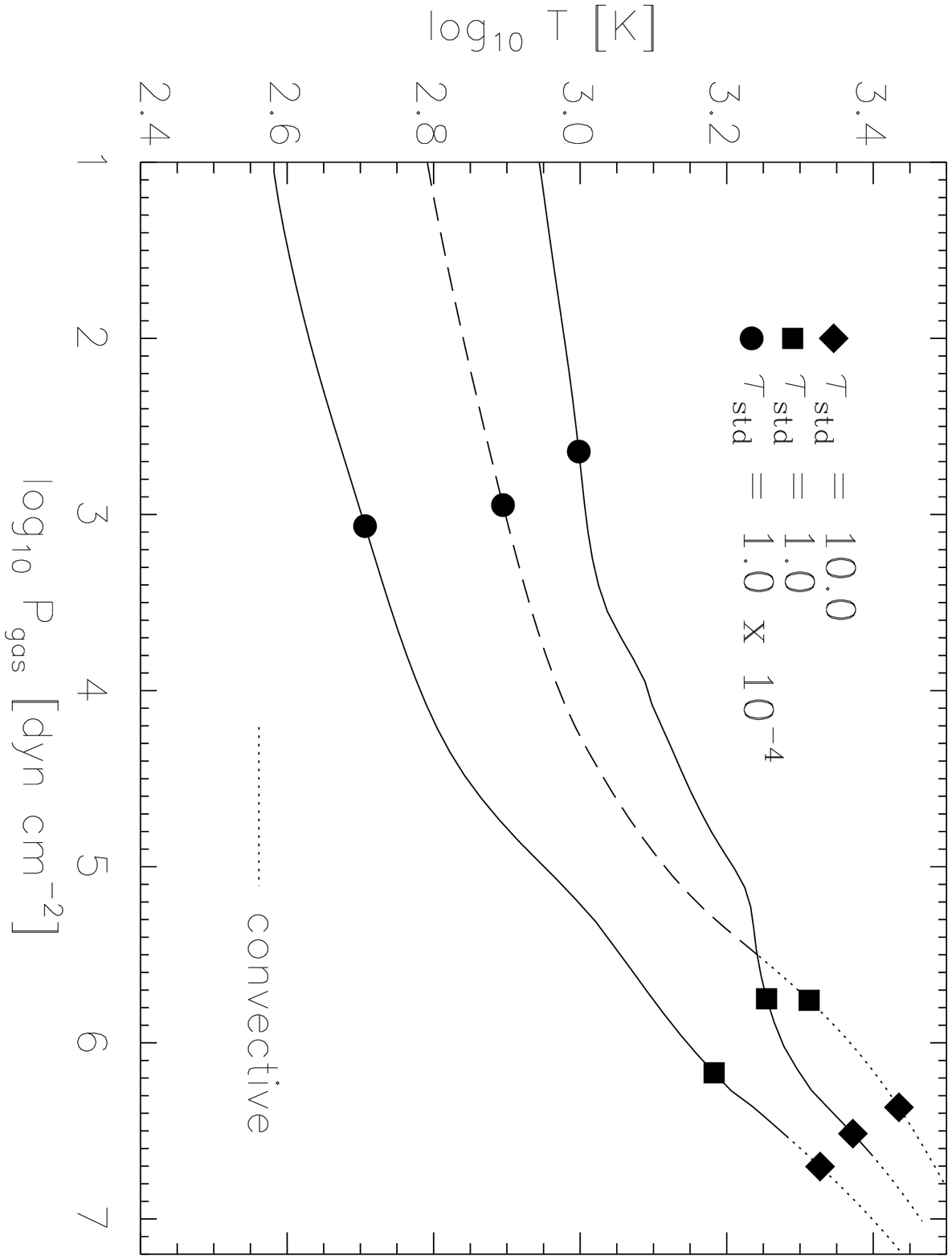]{\label{fig13}The structures  for our 1000$\kel$   AMES-Cond
planet located  0.065AU from a solar  type primary  (top solid line) is
compared to  a non-irradiated structure   of an AMES-Cond  planet with
$\teff = \teq = 1560\kel$ (dashed line). The errors which result from
assuming a non-irradiated  structure are quite  apparent  at {\it all}
depths. For comparison, the non-irradiated 1000$\kel$ AMES-Cond planet
is also  shown   (bottom solid  line).  The filled    symbols refer to
different optical depths ($\tau$) at $\lambda = 1.2\micron$.}

\figcaption[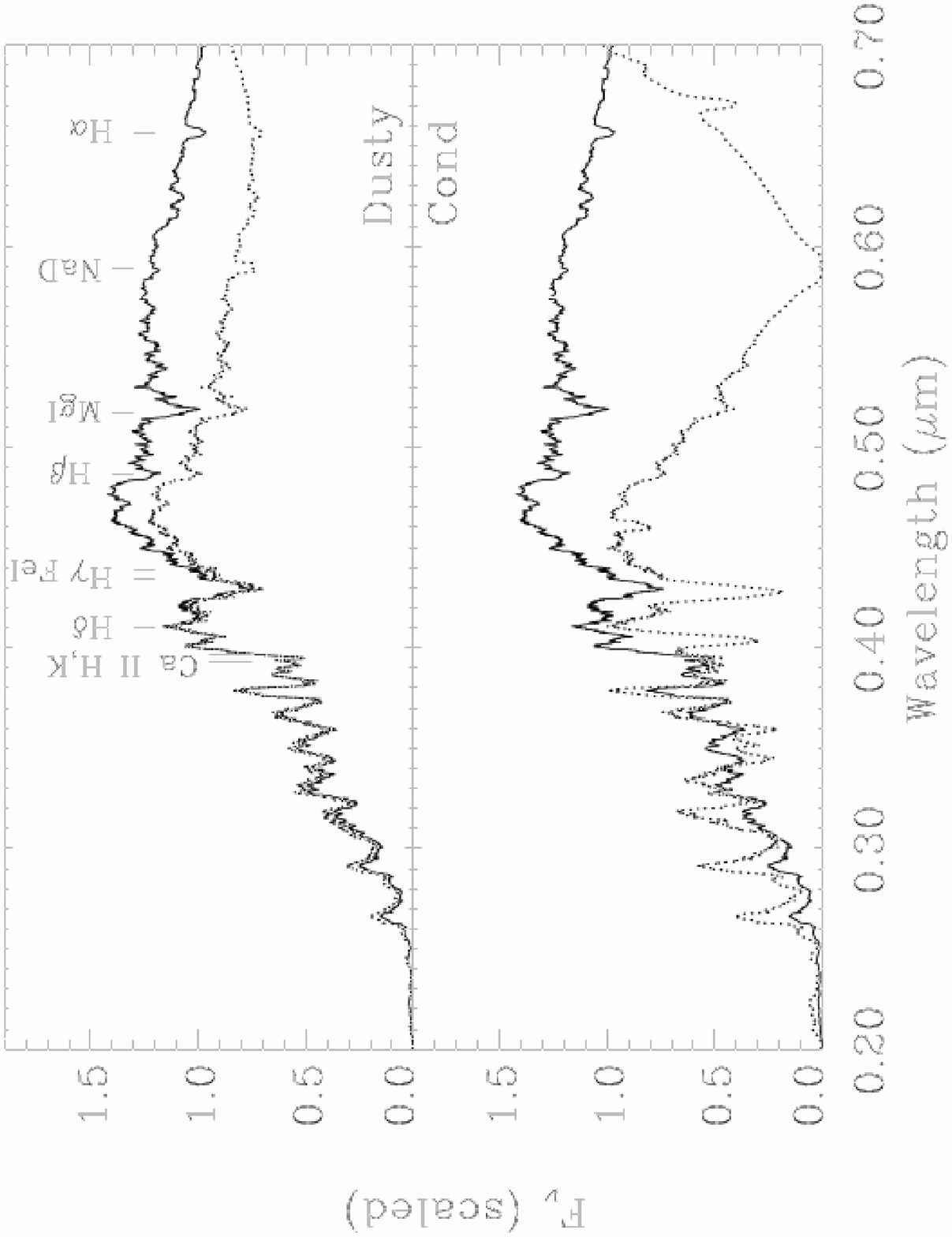]{\label{fig14} The optical spectrum  of a $\tint = 1000\kel$
planet (dotted line) located  at 0.15AU from a  G2 primary is compared
to the incident stellar spectrum (solid line). The top panel shows the
comparison for a  Dusty atmosphere and the  Cond case is  shown in the
lower panel.   The flux has been arbitrarily  scaled to facilitate the
comparison.}

\figcaption[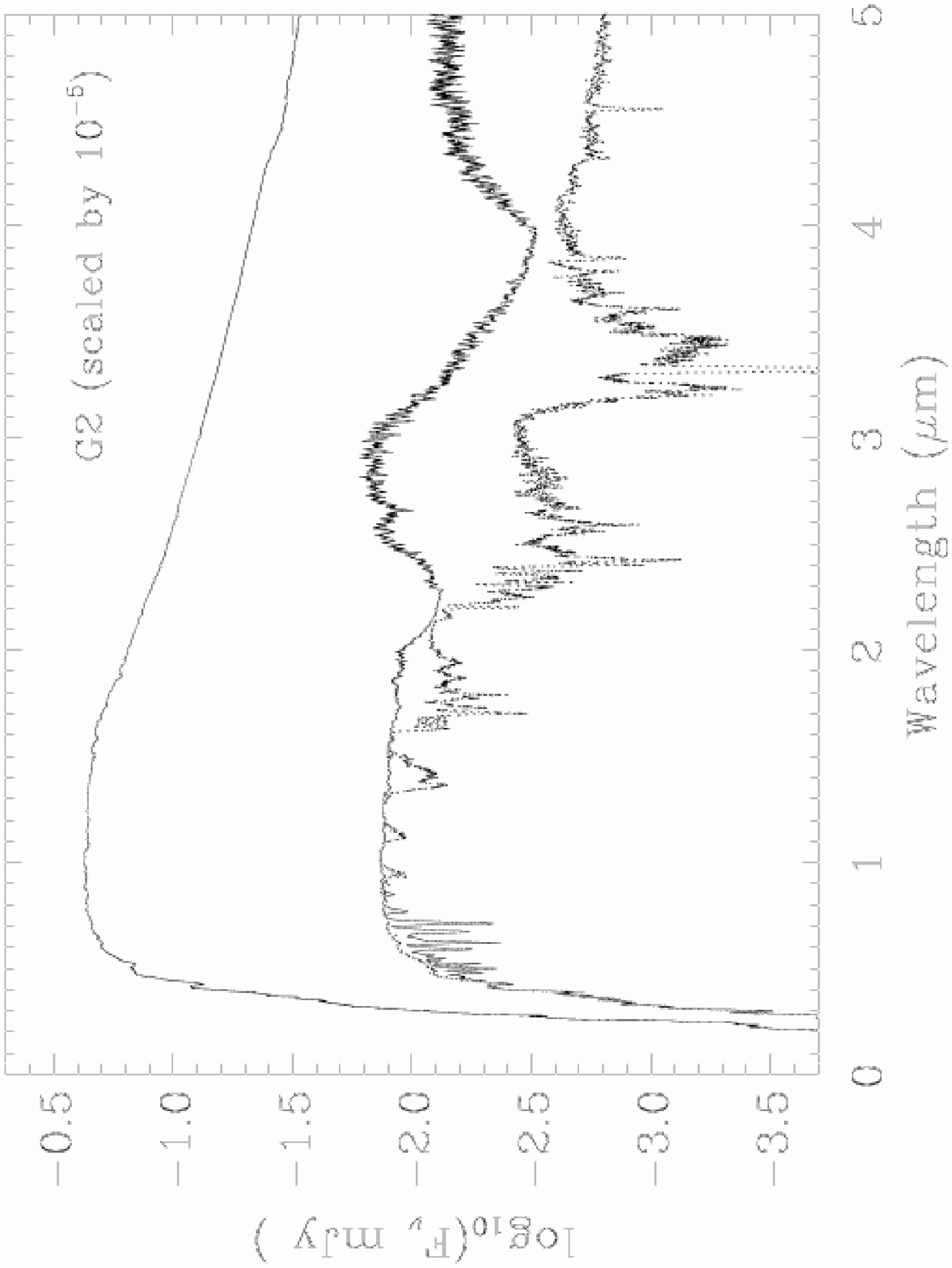]{\label{fig15} The  transmitted flux  through the limb  of a
$1000\kel$  AMES-Cond planet located  0.05AU from  a G2  primary.  The
second  spectrum(full  line)  is  the  transmitted  flux  through  our
irradiated  atmosphere.   The  bottom  spectrum(dotted  line)  is  the
transmitted flux  through our non-irradiated  atmosphere.  All spectra
have been scaled  to 15pc.  For comparison, the spectrum  of a G2 (top
line  scaled  by  an  additional  $10^{-4}$) is  also  shown.  If  the
deposited energy  is not globally redistributed  on short time-scales,
then the transmitted flux should be closer to the non-irradiated case.
Reality is likely to be somewhere in between.}


\begin{thebibliography}{37}
\expandafter\ifx\csname natexlab\endcsname\relax\def\natexlab#1{#1}\fi

\bibitem[{{Allard} {et~al.}(2001){Allard}, {Hauschildt}, {Alexander},
  {Tamanai}, \& {Schweitzer}}]{AMES-2001}
{Allard}, F., {Hauschildt}, P.~H., {Alexander}, D.~R., {Tamanai}, A., \&
  {Schweitzer}, A. 2001, \apj, submitted

\bibitem[{{Allard} {et~al.}(2000){Allard}, {Hauschildt}, \&
  {Schwenke}}]{Allard2000}
{Allard}, F., {Hauschildt}, P.~H., \& {Schwenke}, D. 2000, \apj,
  540, 1005

\bibitem[{Borysow \& Frommhold(1986{\natexlab{a}})}]{n2n2a}
Borysow, A. \& Frommhold, L. 1986{\natexlab{a}}, ApJ, 311, 1043

\bibitem[{Borysow \& Frommhold(1986{\natexlab{b}})}]{h2n2}
---. 1986{\natexlab{b}}, ApJ, 303, 495

\bibitem[{Borysow \& Frommhold(1986{\natexlab{c}})}]{h2ch4}
---. 1986{\natexlab{c}}, ApJ, 304, 849

\bibitem[{Borysow \& Frommhold(1987{\natexlab{a}})}]{n2n2b}
---. 1987{\natexlab{a}}, A\&A, 320, 437

\bibitem[{Borysow \& Frommhold(1987{\natexlab{b}})}]{ch4ch4}
---. 1987{\natexlab{b}}, ApJ, 318, 940

\bibitem[{Borysow {et~al.}(1997{\natexlab{a}})Borysow, Jorgensen, \&
  Cheng}]{h2h2a}
Borysow, A., Jorgensen, U.~G., \& Cheng. 1997{\natexlab{a}}, A\&A, 324, 185

\bibitem[{Borysow {et~al.}(1997{\natexlab{b}})Borysow, Jorgensen, \&
  Cheng}]{h2hea}
---. 1997{\natexlab{b}}, A\&A, 324, 185

\bibitem[{Borysow \& Tang(1993)}]{n2ch4}
Borysow, A. \& Tang. 1993, Icarus, 105, 175

\bibitem[{{Brett} \& {Smith}(1993)}]{Brett93}
{Brett}, J.~M. \& {Smith}, R.~C. 1993, \mnras, 264, 641+

\bibitem[{{Cameron} {et~al.}(1999){Cameron}, {Horne}, {Penny}, \&
  {James}}]{Cameron1999}
{Cameron}, A., {Horne}, K., {Penny}, A., \& {James}, D. 1999, \nat, 402, 751

\bibitem[{{Chabrier} {et~al.}(2000){Chabrier}, {Baraffe}, {Allard}, \&
  {Hauschildt}}]{Chabrier2000}
{Chabrier}, G., {Baraffe}, I., {Allard}, F., \& {Hauschildt}, P. 2000, \apj,
  542, 464

\bibitem[{{Charbonneau} {et~al.}(2000){Charbonneau}, {Brown}, {Latham}, \&
  {Mayor}}]{Charbon00}
{Charbonneau}, D., {Brown}, T.~M., {Latham}, D.~W., \& {Mayor}, M. 2000, \apjl,
  529, L45

\bibitem[{{Charbonneau} {et~al.}(1999){Charbonneau}, {Noyes}, {Korzennik},
  {Nisenson}, {Jha}, {Vogt}, \& {Kibrick}}]{Charbon1999}
{Charbonneau}, D., {Noyes}, R.~W., {Korzennik}, S.~G., {Nisenson}, P., {Jha},
  S., {Vogt}, S.~S., \& {Kibrick}, R.~I. 1999, \apjl, 522, L145

\bibitem[{Chase {et~al.}(1985)Chase, Davis, Downey, Frurip, McDonald, \&
  Syverud}]{janaf}
Chase, M.~W., Davis, C.~A., Downey, J.~R., Frurip, D.~J., McDonald, R.~A., \&
  Syverud, A.~N. 1985, J. Phys. Chem. Ref. Data, 14, Sup.1

\bibitem[{{Delfosse} {et~al.}(1999){Delfosse}, {Forveille}, {Beuzit}, {Udry},
  {Mayor}, \& {Perrier}}]{Delfosse1999}
{Delfosse}, X., {Forveille}, T., {Beuzit}, J.~., {Udry}, S., {Mayor}, M., \&
  {Perrier}, C. 1999, \aap, 344, 897

\bibitem[{{Encrenaz}(1999)}]{Encrenaz99}
{Encrenaz}, T. 1999, \aapr, 9, 171

\bibitem[{{Goukenleuque} {et~al.}(2000){Goukenleuque}, {B{\'e}zard}, {Joguet},
  {Lellouch}, \& {Freedman}}]{Gouken2000}
{Goukenleuque}, C., {B{\'e}zard}, B., {Joguet}, B., {Lellouch}, E., \&
  {Freedman}, R. 2000, Icarus, 143, 308

\bibitem[{{Grossman}(1972)}]{Grossman72}
{Grossman}, L. 1972, \gca, 38, 47

\bibitem[{Gruszka \& Borysow(1997)}]{co2co2}
Gruszka, M. \& Borysow, A. 1997, Icarus, 129, 172

\bibitem[{{Guillot}(2000)}]{Guillot00}
{Guillot}, T. 2000, in IAU Symposium, Vol. 202, E19--+

\bibitem[{{Guillot} {et~al.}(1996){Guillot}, {Burrows}, {Hubbard}, {Lunine}, \&
  {Saumon}}]{Guillot96}
{Guillot}, T., {Burrows}, A., {Hubbard}, W.~B., {Lunine}, J.~I., \& {Saumon},
  D. 1996, \apjl, 459, L35

\bibitem[{{Hauschildt} {et~al.}(1999){Hauschildt}, {Allard}, \&
  {Baron}}]{Nextgen99}
{Hauschildt}, P.~H., {Allard}, F., \& {Baron}, E. 1999, \apj, 512, 377

\bibitem[{Hauschildt \& Baron(1999)}]{jcam}
Hauschildt, P.~H. \& Baron, E. 1999, Journal of Computational and Applied
  Mathematics, 102, 41

\bibitem[{{Henry} {et~al.}(2000){Henry}, {Marcy}, {Butler}, \&
  {Vogt}}]{Henry00}
{Henry}, G.~W., {Marcy}, G.~W., {Butler}, R.~P., \& {Vogt}, S.~S. 2000, \apjl,
  529, L41

\bibitem[{{Henry}(1998)}]{Henry1998}
{Henry}, T.~J. 1998, in ASP Conf. Ser. 134: Brown Dwarfs and Extrasolar
  Planets, 28+

\bibitem[{Husson {et~al.}(1992)Husson, Bonnet, Scott, \& A.}]{geisa}
Husson, N., Bonnet, B., Scott, N., \& A., C. 1992, JQSRT, 48, 509

\bibitem[{{Leggett} {et~al.}(2000){Leggett}, {Allard}, {Dahn}, {Hauschildt},
  {Kerr}, \& {Rayner}}]{Leggett2000}
{Leggett}, S.~K., {Allard}, F., {Dahn}, C., {Hauschildt}, P.~H., {Kerr}, T.~H.,
  \& {Rayner}, J. 2000, \apj, 535, 965

\bibitem[{{Marcy} {et~al.}(1998){Marcy}, {Butler}, {Vogt}, {Fischer}, \&
  {Lissauer}}]{Marcy1998}
{Marcy}, G.~W., {Butler}, R.~P., {Vogt}, S.~S., {Fischer}, D., \& {Lissauer},
  J.~J. 1998, \apjl, 505, L147

\bibitem[{{Marley} {et~al.}(1999){Marley}, {Gelino}, {Stephens}, {Lunine}, \&
  {Freedman}}]{Marley1999}
{Marley}, M.~S., {Gelino}, C., {Stephens}, D., {Lunine}, J.~I., \& {Freedman},
  R. 1999, \apj, 513, 879

\bibitem[{Partridge \& Schwenke(1997)}]{ames-water-new}
Partridge, H. \& Schwenke, D.~W. 1997, J. Chem. Phys., 106, 4618

\bibitem[{Rothman {et~al.}(1992)Rothman, Gamache, Tipping, Rinsland, Smith,
  {Chris Benner}, {Malathy Devi}, Flaub, Camy-Peyret, Perrin, Goldman, Massie,
  Brown, \& Toth}]{hitran92}
Rothman, L.~S., Gamache, R.~R., Tipping, R.~H., Rinsland, C.~P., Smith, M.
  A.~H., {Chris Benner}, D., {Malathy Devi}, V., Flaub, J.-M., Camy-Peyret, C.,
  Perrin, A., Goldman, A., Massie, S.~T., Brown, L., \& Toth, R.~A. 1992,
  JQSRT, 48, 469

\bibitem[{Samuelson {et~al.}(1997)Samuelson, Nath, \& Borysow}]{ch4ar}
Samuelson, R.~E., Nath, N., \& Borysow, A. 1997, Planetary \& Space Sciences,
  45/8, 959

\bibitem[{{Saumon} {et~al.}(1996){Saumon}, {Hubbard}, {Burrows}, {Guillot},
  {Lunine}, \& {Chabrier}}]{Saumon1996}
{Saumon}, D., {Hubbard}, W.~B., {Burrows}, A., {Guillot}, T., {Lunine}, J.~I.,
  \& {Chabrier}, G. 1996, \apj, 460, 993+

\bibitem[{{Seager} \& {Sasselov}(1998)}]{Seager1998}
{Seager}, S. \& {Sasselov}, D.~D. 1998, \apjl, 502, L157

\bibitem[{{Seager} \& {Sasselov}(2000)}]{Seager2000a}
---. 2000, \apj, 537, 916

\bibitem[{{Seager} {et~al.}(2000){Seager}, {Whitney}, \&
  {Sasselov}}]{Seager2000b}
{Seager}, S., {Whitney}, B.~A., \& {Sasselov}, D.~D. 2000, \apj, 540, 504

\bibitem[{{Sudarsky} {et~al.}(2000){Sudarsky}, {Burrows}, \&
  {Pinto}}]{Sudarsky2000}
{Sudarsky}, D., {Burrows}, A., \& {Pinto}, P. 2000, \apj, 538, 885

\end{thebibliography}
\end{document}